# EIGENRAYS IN 3D HETEROGENEOUS ANISOTROPIC MEDIA: PART VII – DYNAMICS, FINITE-ELEMENT IMPLEMENTATION

*Igor Ravve (corresponding author) and Zvi Koren, Emerson*

*igor.ravve@emerson.com , zvi.koren@emerson.com*

## ABSTRACT

In this part, we apply the same finite-element approach, used in Part III for the vanishing first traveltime variation (to obtain the stationary rays), for the second traveltime variation, in order to compute the dynamic characteristics along the stationary ray. The finite-element solver involves application of the weak formulation and the Galerkin method to the linear second-order Jacobi ordinary differential equation (derived in Part V), yielding an original linear algebraic equation set for dynamic ray tracing. In our formulation, the resolving matrix of the linear equation set coincides with the global traveltime Hessian computed for the kinematic ray tracing, making the solution of the dynamic problem straightforward. The proposed method is unconditionally stable (the solution does not explode when the intervals between the nodes are increased) and is more accurate than the commonly used numerical integration (e.g., Runge-Kutta) methods, in particular, for stationary rays passing through heterogeneous anisotropic models with complex wave phenomena.

Keywords: General anisotropy, Finite element method, Paraxial rays, Geometric spreading, Caustics, KMAH index.

## INTRODUCTION

The proposed Eigenray method is a finite-element approach, primarily designed to solve the kinematic (Parts I, II and III) and dynamic (Parts IV, V, VI and VII) two-point ray tracing problem in 3D heterogeneous anisotropic media. In particular, in Part IV we present an efficient (but limited) method to compute the total geometric spreading by compressing the already computed traveltime Hessian matrix into an endpoint traveltime Hessian (related to the source and receiver locations). Parts V, VI and VII are devoted to the computation of dynamic properties along the stationary ray path using the so called, dynamic ray tracing (DRT). In Part V, we provide a comprehensive review of the DRT studies and derive the proposed Jacobi DRT ordinary differential equation to be solved, following Bliss (1916). In Part VI, we compare the Lagrangian and Hamiltonian approaches to the DRT and derive the relationships between the corresponding Lagrangian's and Hamiltonian's Hessian matrices.

In this part (Part VII), we solve the Jacobi DRT using the same finite-element discretization and Hermite interpolation used in Part III for obtaining the stationary rays. We apply the weak formulation and the Galerkin method to the Jacobi DRT equation obtained in Part V (with arbitrary initial or boundary conditions), resulting in a local, first-order, weighted residual, linear algebraic equation set. This set includes matrix blocks which are the local "stiffness" matrices for (either the two-node or three-node) Hermite-type finite elements used in this study. We show that the derived element "stiffness" matrices coincide with the corresponding local traveltime Hessian matrices obtained in Part III, and thus the global traveltime Hessian matrices (after the assembly of the individual element matrices into the whole ray path matrix) are also identical. Although not a surprise, this is a remarkable result that makes it possible to use the already computed traveltime Hessian matrices (used for the kinematic solution) for the computation of the dynamic properties as well.

To demonstrate the implementation of the proposed method, we use several benchmark models: two constant gradient velocity models (with vertical and tilted gradients), a conic velocity model (Ravve and Koren, 2007), a simple caustic-generating model, and a gas-cloud caustic-generating model (Brandsberg-Dahl et al., 2003). We swap the source and receiver in the tested models to validate the reciprocity characteristic of the relative geometric spreading. We then demonstrate the accuracy of the computed kinematic and dynamic properties of a ray propagating in the vertical symmetry plane of an inhomogeneous elliptic orthorhombic model with a tilted reference velocity gradient, considering two cases: an elliptic trajectory for the constant gradient, and an asymmetric trajectory for the spatially varying gradient. The latter case, with the different ratios between the ray and phase velocities at the source and receiver, is important for validating the relationship between the ray Jacobian and the relative geometric spreading in anisotropic media. For all the examples, we first compute the stationary ray paths, and then compute the geometric spreading and analyze these trajectories for possible caustics. Our primary aim is to emphasize the advantages, transparency and simplicity of the suggested approach.

Appendices

In order to make the paper more readable, the body of the paper only contains the main concepts of the finite-element implementation of the proposed Lagrangian-based Jacobi DRT approach, with the principal governing equations and numerical examples, with minimum mathematical derivations. The detailed derivations have been moved to the appendices.

In Appendix A, we apply the weak, weighted-residual, finite-element formulation and the Galerkin method to the Jacobi DRT equation, in order to obtain the finite-element solver in the form of a linear algebraic equation set. We show that the matrix of this solver coincides with the already computed traveltime Hessian matrix for the stationary ray path, making our method efficient.

The rest of the appendices are related to the numerical examples.

In Appendix B, we explain the theory of diving rays for a constant vertical or tilted velocity gradient model in isotropic media.

In Appendix C, the diving waves are explained for the conic velocity model.

Appendices D, E and F are devoted to the so-called simple caustic-generating model. In Appendix D we describe the theory of diving rays in this model, consisting of a constant velocity layer and a constant velocity gradient half-space. In Appendix E we compute analytically the Jacobian for any diving ray in this model, with and without caustics. In Appendix F, we present a ray in such media using a function with only two parameters (two DoF) and demonstrate analytically the existence of a saddle-point stationary path.

In Appendix G, the theory of diving rays is extended for the factorized inhomogeneous anisotropic (FIA) media (an ellipsoidal orthorhombic symmetry) with a tilted gradient of the reference velocity.

## JACOBI DRT SET AND ITS FINITE-ELEMENT SOLVER

In this section, we derive the weak finite-element formulation for the Jacobi DRT equation, and we obtain the linear algebraic DRT solver.

The Jacobi DRT equation obtained in Part V (based on the fundamental study by Bliss, 1916) reads,

$$\frac{d}{ds}\left(L_{\mathbf{rx}} \cdot \mathbf{u} + L_{\mathbf{rr}} \cdot \dot{\mathbf{u}}\right) = L_{\mathbf{xx}} \cdot \mathbf{u} + L_{\mathbf{xr}} \cdot \dot{\mathbf{u}} \quad . \quad (1)$$

It contains the second derivative of the normal shift $\mathbf{u}$ with respect to (wrt) the arclength of the central ray, $\ddot{\mathbf{u}}$. In order to eliminate this second derivative, we apply the weak formulation to this equation, locally, for each finite element. This means multiplying the vector-form ODE (equation 1) by a set of scalar weight functions (one function at a time), and integrate over the finite-element arclength,

$$\int_{s_{\text{ini}}}^{s_{\text{fin}}} \frac{d}{ds}\left(L_{\mathbf{rx}} \cdot \mathbf{u} + L_{\mathbf{rr}} \cdot \dot{\mathbf{u}}\right) w(s) ds = \int_{s_{\text{ini}}}^{s_{\text{fin}}} \left(L_{\mathbf{xx}} \cdot \mathbf{u} + L_{\mathbf{xr}} \cdot \dot{\mathbf{u}}\right) w(s) ds \quad , \quad (2)$$

where $s_{\text{ini}}$ and $s_{\text{fin}}$ are the values of the arclength at the endpoints of a single finite element, $s_{\text{ini}} < s_{\text{fin}}$. The Galerkin method assumes that the weight (test) functions $w(s)$ are the same as the interpolation functions within a finite element; in our case, these are Hermite interpolation polynomials. According to the Galerkin (1915) method, the residual of the differential equation is orthogonal to each of the test functions. This effectively reduces the second-order ODE set to the first-order, local, weighted residual, linear algebraic equation set,

$$\int_{\xi=-1}^{\xi=+1} \left(L_{\mathbf{rx}} \cdot \mathbf{u} + L_{\mathbf{rr}} \cdot \dot{\mathbf{u}}\right) \frac{dw}{d\xi} d\xi + \int_{\xi=-1}^{\xi=+1} \left(L_{\mathbf{xx}} \cdot \mathbf{u} + L_{\mathbf{xr}} \cdot \dot{\mathbf{u}}\right) w \frac{ds}{d\xi} d\xi = \left(L_{\mathbf{rx}} \cdot \mathbf{u} + L_{\mathbf{rr}} \cdot \dot{\mathbf{u}}\right) w \Big|_{\xi=-1}^{\xi=+1} \quad , \quad (3)$$

where $-1 \leq \xi \leq +1$ is the internal flow parameter within a single finite element, $s = s(\xi)$, $s(-1) = s_{\text{ini}}$, $s(+1) = s_{\text{fin}}$. The arclength of the central ray is related to the internal flow parameter my means of the metric, $ds/d\xi$. Note that equation 3 does not include the second derivative of the shift, $\ddot{\mathbf{u}}(s)$. The nodal values of the solution and its derivative are so far unknown, and Hermite interpolation is applied between the nodes.

The weak formulation includes integration by parts, which, in turn, yields the boundary terms on the right side of equation 3. These boundary terms at the end nodes of the joined elements cancel each other at the assembly, due to their equal values of opposite signs. Eventually, only the boundary terms of the source and receiver remain; these terms correspond to the (arbitrary) initial or boundary conditions of a paraxial ray at the endpoints of the path.

Assembling the element matrices into the global matrix of the whole path and taking into account the constraints discussed in Part V,

$$\mathbf{u} \cdot \mathbf{r} = 0 \quad \text{and} \quad \frac{d}{ds}(\mathbf{u} \cdot \mathbf{r}) = \dot{\mathbf{u}} \cdot \mathbf{r} + \mathbf{u} \cdot \dot{\mathbf{r}} = 0 \qquad , \qquad (4)$$

we obtain the final linear finite-element solver for the vector-form Jacobi DRT equation, with arbitrary initial or boundary conditions (see Appendix A for details). Applying the solver, we obtain the nodal values of the normal shift $\mathbf{u}(s)$ and its derivative $\dot{\mathbf{u}}(s)$ wrt the arclength of the central ray for any specified initial or boundary conditions. The Hermite interpolation provides the values of these functions along intervals between the nodes

## NUMERICAL EXAMPLES

The following numerical examples can be considered benchmark problems, where the objective is mainly to validate the theory presented in this work. More realistic models with different anisotropic symmetries will be the target of our next study. In this part of the study, we present five numerical examples of the kinematics (stationary rays) and dynamics (geometric spreading with caustic location and classification) computed with the Eigenray method for: a constant velocity gradient model, a conic model, and two caustic-generating models. The last example presents a diving ray in an ellipsoidal anisotropic model with tilted reference velocity gradient. In the examples below, eight three-node finite elements were used to present the ray path. Obviously, for real field examples, the number of nodes is much higher.

Example 1: Geometric spreading for diving rays in media with constant velocity gradients.

Example 1a. A constant vertical velocity gradient model is defined by two parameters: surface velocity $v_a$ and vertical gradient $k$. We assume that $h = 10$ km, $v_a = 2$ km/s, $k = 1$ s$^{-1}$, where $h$ is the surface offset (the chord of the circular arc). We solve the problem for the stationary path with eight three-node elements (see Figure 13 of Part III with a similar scheme but for three elements), and obtain the nodal locations and orientations shown in Table 1 and Figure 1a (a solid line, where colors correspond to different finite elements). The initial guess is the straight line on the surface, connecting the source and receiver. The finite-element traveltime coincides with the theoretical value up to eleven digits. The geometric spreading $L_{GS}$ along the stationary ray path and the normalized geometric spreading $L_{GS}/\sigma$ are plotted by solid lines in Figures 1b and 1c, respectively. As we expected, $L_{GS} = \sigma$ for the constant velocity gradient model. The condensed $6 \times 6$ source-receiver Hessian $\nabla_{SR}\nabla_{SR}t$ is presented in Table 2, where its mixed $3 \times 3$ block is highlighted in yellow. Only the upper left $2 \times 2$ sub-matrix of the highlighted block is used. This sub-matrix is applied to obtain geometric spreading for the whole path (i.e., between source and receiver), and the result is identical to the geometric spreading obtained with the Jacobi DRT solution at the receiver point $R$.

Applying the relationships derived in Appendix B, we obtain the theoretical values of the ray path parameters: take-off angle, radius of trajectory, traveltime, geometric spreading, etc. In Table 3 we list the values of these parameters for the stationary ray path and the accuracy of their numerical computation. In this example, the accuracy of the geometric spreading obtained using the finite element method is excellent, although the accuracy of the traveltime is better. This is not a surprise: The accuracies of the second derivatives of a function are normally worse than the accuracy of the function itself.

Example 1b. Next, we test equation 28 of part V for the conversion velocity that relates the ray Jacobian to the (relative) geometric spreading. This equation includes the ray velocity at the source and does not include it at the receiver (or at a current ray path point). In this test, we consider a constant tilted gradient model with the midpoint velocity $v_a = 3\,\text{km/s}$, the gradient components $k_1 = 0.2\,\text{s}^{-1}$, $k_3 = 0.8\,\text{s}^{-1}$, and the offset $h = 10\,\text{km}$. The source and receiver are located at $x_1 = \mp h/2$, respectively, and the medium velocities at these points are different: $v_S = 2\,\text{km/s}$, $v_R = 4\,\text{km/s}$. Applying the same finite-element scheme as in the previous case, we compute the ray path (that proves to be symmetric about the vertical midpoint line, despite the lateral gradient component) and its kinematic and dynamic characteristics, with an excellent accuracy. Appendix B makes it possible to establish these parameters analytically. The traveltime proves to be accurate up to eleven digits. The ray path is presented in Table 4 and Figure 2a, the source-receiver Hessian – in Table 5, the numerical path characteristics, along with their error values, – in Table 6. Figure 2b shows the geometric spreading along the ray (a solid line). The normalized geometric spreading is identically 1 along the ray path, as shown in Figure 2c. We then swap the source and receiver and obtain the same results. The geometric spreading vs. the arclength along the ray path is, of course, different, as shown by dashed line in Figure 2b, but its values at the final point of the ray trajectory are identical for the forward and reverse paths, $L_{GS} = 50\,\text{km}^2/\text{s}$.

Example 2: Geometric spreading for a diving ray in a medium with a conic velocity model

The conic velocity model is described by three parameters: surface velocity $v_a$, surface gradient $k_a$, and an additional asymptotic (bounding) velocity $v_\infty$. We assume the same data/model as in the previous example for the constant vertical velocity gradient, $h = 10\,\text{km}, v_a = 2\,\text{km/s}, k_a = 1\,\text{s}^{-1}$, and the asymptotic velocity is set to $v_\infty = 6\,\text{km/s}$. The conic velocity profile and its gradient profile are plotted in Figures 3a and 3b, respectively. The diving rays in a medium with a conic velocity model are explained in Appendix C. Applying this theory, we obtain the theoretical values for horizontal slowness $p_h$, eccentricity $m$, take-off angle $\theta_a$, parameter $\sigma$, and the major and minor semi-axes $A_e, B_e$ of the elliptic arc, respectively,

$$\begin{aligned} p_h &= 0.259195\,\text{s/km}, & m &= 0.643016, \\ \theta_a &= 0.544967\,\text{rad}, & \sigma &= 38.5810\,\text{km}^2/\text{s}, \\ A_e &= 5.51257\,\text{km}, & B_e &= 4.22182\,\text{km}. \end{aligned} \quad (5)$$

We apply the same finite-element scheme and the same initial guess as in the previous example. The nodal locations and orientations are listed in Table 7. The stationary ray path, the non-normalized and the normalized geometric spreading are plotted by dashed lines in Figures 1a, 1b and 1c, respectively. Again, the finite-element traveltime coincides with the theoretical value up to eleven digits. The condensed, source-receiver traveltime Hessian, used for alternative computation of the geometric spreading for the whole path, is presented in Table 8, with the mixed block highlighted in yellow. The geometric spreading computed with this block is identical to the Jacobi solution at the receiver. The elliptic ray path parameters and the accuracy of their computation are presented in Table 9.

Note that for the constant gradient model, velocity vs. depth is higher, the maximum penetration $z_{max}$ of the diving ray is deeper, the arclength is longer, and the traveltime is shorter than those for the conic velocity model. As we see in Figure 1a, the maximum depth of the circular path (constant gradient model) exceeds that of the elliptic (conic model). The reason is that the velocities for the linear model are higher at the same depths. For the same reason, the traveltime of the circular path is shorter. Note that at the receiver point, the normalized geometric spreading for the conic velocity model, $L_{GS}/\sigma = 1.42580$, which means that one should be careful when using $\sigma$ as an approximation for $L_{GS}$ even in simple velocity models.

Example 3: Geometric spreading for a diving ray in a simple caustic-generating velocity model.

Caustics may occur in layered media with discontinuous increase of the velocity gradient (e.g., Murphy, 1961; Bott, 1982; Nye, 1985; Cygan, 2006; Aster, 2011, and many others). The simplest presentation of such a medium is a constant velocity layer over a constant velocity gradient half-space. We assume the following parameters: the layer thickness, $z_h = 3\,\text{km}$, its velocity, $v_a = 1\,\text{km/s}$, and the half-space velocity gradient, $k_o = 1\,\text{s}^{-1}$. The velocity is continuous at the interface, but the gradient is not, and the second derivative of the velocity is singular (representing a delta-function). To apply the Eigenray approach, we smooth the velocity profile along the vertical axis $z$ and apply the following function for the vertical velocity gradient,

$$\frac{dv}{dz} = k(z) = \frac{k_o}{2}\left(1 + \tanh\frac{z - z_h}{\Delta z_h}\right) \quad , \tag{6}$$

where $z \equiv x_3$, parameter $z_h$ is the constant-velocity layer thickness, and the vertical width parameter $\Delta z_h = 0.1 \text{ km}$ is responsible for the smoothness. The thickness of the gradient transition zone is approximately $5\Delta z_h$. The velocity profile is obtained by integrating the gradient in depth, with the initial condition,

$$\lim_{z \to -\infty} v(z) = v_a \quad . \tag{7}$$

This leads to,

$$v(z) = v_a + \frac{k_o(z - z_h)}{2} + \frac{k_o \Delta z_h}{2} \ln\left(2\cosh\frac{z - z_h}{\Delta z_h}\right) \quad . \tag{8}$$

The second derivative of the velocity then reads,

$$\frac{d^2v(z)}{dz^2} = \frac{dk(z)}{dz} = \frac{k_o}{2\Delta z_h} \cosh^{-2}\frac{z - z_h}{\Delta z_h} \quad . \tag{9}$$

The maximum value of the second derivative occurs at the centerline of the transition zone, $z = z_h$, and is equal to $k_o/(2\Delta z_h)$. The velocity profile, velocity gradient, and second derivative of the velocity are shown in Figure 4. The kinematic characteristics of diving rays propagating in this velocity model are explained in Appendix D, and dynamic characteristics (i.e., computation of the Jacobian for the transform between the Cartesian and ray coordinates) are provided in Appendix E. In a medium with this velocity model, a diving ray is possible only for offsets exceeding a definite threshold minimum, $h_{min}$. For any offset exceeding the minimum value, two diving rays co-exist (multi-arrivals). The ray with a smaller take-off angle is caustic-free, while the one with a larger take-off angle exerts a caustic. For the minimum-offset ray, a single diving ray exists, with a caustic located exactly at the destination point.

The rays in the simple caustic-generating model can be classified into three types: a) pre-critical caustic-free rays with the take-off angle $\theta_a < \theta_c$, b) a critical ray with $\theta_a = \theta_c$ and a caustic located exactly at the receiver, and c) post-critical rays with $\theta_a > \theta_c$ and a caustic located at an internal point of the path. The critical angle $\theta_c$ in an unsmooth simple caustic-generating model is defined by the model parameters,

$$\theta_c = \arctan \sqrt{\Delta z_v / z_h} \qquad . \qquad (10)$$

The ray paths computed in the simple caustic-generating medium are shown in Figures 5a (pre-critical take-off angle), 6a (post-critical angle) and 7a (critical angle). Eight three-node finite elements were used (17 nodes, 96 internal DoF), and the corresponding segments of the paths are shown by different colors. For each case, the gray dashed line shows an initial guess. For the pre-critical and critical rays, the initial paths are elliptic arcs, given the offset, take-off angle, and maximum depth. For the post-critical ray, the initial path is a hyperbole, given the offset, maximum depth, and radius of curvature at the apex. As shown in Figures 5a, 6a and 7a, in all three cases the maximum depth of the initial paths was deliberately over-estimated. Applying the theory of Appendix D, we computed analytically the stationary path parameters and listed them in Table 10. The parameters computed numerically with the finite element method are listed in Table 11. They are very close to the analytical parameters of Table 10, with minor discrepancies caused by smoothing of the gradient discontinuity in the medium used for the numerical analysis.

It is interesting to note that while the stationary path of the deep diving ray (caustic-free) and the trivial straight-line path connecting the source and the receiver are true traveltime minima, the stationary path of the shallow ray (with the caustic) represents a saddle point of the traveltime. The maximum penetration depth of the diving ray is one of its DoF. A zero depth corresponds to a trivial solution – a straight line between the source and receiver that delivers a minimum traveltime. The deep diving ray of the stationary penetration depth delivers another minimum traveltime. A maximum inevitably exists between these two minima, and parameter $z_{max}$ corresponding to this maximum is the penetration depth of the ray with the high take-off angle and the caustic (shallow diving ray). This ray corresponds to a maximum for one DoF, and minima for all other DoF; hence, its stationary state represents a saddle point. See a discussion in Appendix F for details.

When a ray path includes multiple DoF, it is unlikely to obtain a path of a true traveltime maximum (where all eigenvalues of the traveltime Hessian are negative); such a maximum normally does not exist. However, a saddle point may be the case where a ray path has a minimum time (or a minimum Hamilton action in particle mechanics) wrt some nearby alternative curves and a maximum wrt others (e.g., Gray and Taylor, 2007). For the shallow diving ray with a caustic, we checked the eigenvalues of the traveltime Hessian matrix of the finite-element scheme for the stationary ray, subjected to kinematic boundary conditions, to make sure that this is indeed a saddle point. One of the eigenvalues proved to be negative, and the others are positive.

In some simple benchmark problems for bending reflection rays, with one or two internal DoF only, it is possible to obtain a traveltime maximum wrt these DoF, so that all neighbor non-stationary trial ray paths will have shorter traveltimes. This is, however, not the case for numerical solutions with multiple DoF: It is very unlikely that all eigenvalues of the global traveltime Hessian matrix will be negative.

To compute geometric spreading and to detect caustics, we applied the proposed Jacobi finite-element solver for the three rays mentioned above (pre-critical, post-critical and critical), with offsets of $h = 1.2 h_{min}$ and $h = h_{min}$. The geometric spreading $L_{GS}(s)$, is plotted in Figures 5b, 6b and 7b for the pre-critical, post-critical and critical rays, respectively, where $v_S$ is the velocity at the source. The normalized geometric spreading $L_{GS}/\sigma$ is plotted for the three rays in Figures 5c, 6c and 7c.

The results obtained for rays in the caustic-generating medium are in agreement with the theoretical predictions described in Appendices D and E.

Example 4: Geometric spreading for 2.5D and 3D gas-cloud velocity model

The gas-cloud model was suggested by Brandsberg-Dahl et al. (2003) to test an algorithm of generating Common Image Gathers (CIG) based on the Generalized Radon Transform (GRT). This model represents a simplification of the real geology at the Valhall field, located in the Norwegian sector of the North Sea and studied by O'Brien et al. (1999). We further simplified the model and rounded off the numerical input data. In this example, the gas-cloud model represents either a cylindrical (2.5D) or a spherical (3D) low-velocity anomaly over a background constant velocity gradient half-space, referred to in Examples 4a and 4b, respectively. The velocity field is given by,

$$v = v_a + k\, x_3 - \Delta v \exp\left[-\frac{(x_1 - c_1)^2 + (x_3 - c_3)^2}{\Delta x_o^2}\right] \quad \text{2.5D},$$

$$v = v_a + k\, x_3 - \Delta v \exp\left[-\frac{(x_1 - c_1)^2 + (x_2 - c_2)^2 + (x_3 - c_3)^2}{\Delta x_o^2}\right] \quad \text{3D},$$

(11)

where $v_a = 1.5$ km/s is the background surface velocity, $k = 0.5\,\text{s}^{-1}$ is the background vertical velocity gradient, $\Delta v = 0.8$ km/s is the magnitude of the velocity anomaly (the gas cloud), $\Delta x_o = 0.3$ km is a parameter governing the width of the transition (smoothing) zone, and $\begin{bmatrix} c_1 & c_2 & c_3 \end{bmatrix} = \begin{bmatrix} 0 & 0 & 1\,\text{km} \end{bmatrix}$ is the cloud center location. In this example, we study a one-way path, with a source located at the origin $x_3 = 0$, and a receiver located under the source, at a depth $x_3 = d_o = 3$ km. The velocity distribution and the absolute value of the velocity gradient vector are shown for the 2.5D case in Figures 8a and 8b, respectively. The kinematics of the 2.5D and 3D cases are identical (Table 12), while the dynamics are different (Table 13). In both cases we deal with multi-arrivals: The vertical ray passes through the cloud center and delivers a saddle point traveltime, while the bypassing "lateral" rays deliver the minimum traveltime solutions. The ray path and geometric spreading related to the vertical ray are shown by solid lines in Figures 9 and 10, while those related to the bypassing rays – by dashed lines. There are two symmetric bypassing rays for the 2.5D model, as shown in Figure 9a, and a bypassing ray per any azimuth for the 3D model. In the 3D case, we chose a fixed azimuth of the vertical plane $x_1 x_3$. Figures 9b and 9c show the non-normalized and normalized geometric spreading, respectively, for the vertical ("passing through") and bypassing rays of the 2.5D model. The vertical ray has a caustic at a depth 1.5 km, which does not coincide with the depth of the cloud center, 1 km (a similar result was obtained by Brandsberg-Dahl et al., 2003). Further analysis (equations F3 and F4 of Part V) shows that this is a line (first-order) caustic, where the line is parallel to the axis of the cylindrical cloud, $x_2$. The bypassing rays in the 2.5D model do not have caustics. Figures 10a and 10b show the non-normalized and normalized geometric spreading, respectively, for the vertical and bypassing rays in the 3D model. The vertical ray has a caustic at the same depth 1.5 km, which is 0.5 km below the center of the spherical cloud, but now equation F4 of Part V shows that this is a point caustic. The bypassing rays have a caustic at the destination (receiver) point $d_o = 3$ km, and equations F3 and F4 of Part V show that this is a line caustic, where the line direction belongs to the plane of the curvilinear ray trajectory and is normal to the ray (see arrival angle $\theta_b$ in Table 12). The physical nature of the line caustic at the receiver differs from that of the point caustic at the depth 1.5 km. The endpoint (receiver) caustic is not directly related to the cloud. The 3D endpoint caustic exists due to the multiplicity of the bypassing ray trajectories for the fixed source and receiver, which in turn, is a sequence of the radial symmetry. A solution exists for any azimuth, and a paraxial ray of a slightly different azimuth yields the same traveltime. Therefore, despite a rule that a receiver caustic (located exactly at the end of the ray trajectory) is a limit case between a minimum time and a saddle point time, we can still consider the bypassing ray trajectory in the 3D model as a minimum time path, just keeping in mind that all azimuths deliver the same minimum traveltime.

We note that for this specific model, the derivative of the geometric spreading wrt the arclength is infinite for the point (first-order) caustic (solid lines in Figures 9b and 9c, and dashed lines in Figures 10a and 10b), and finite for the point (second-order) caustic (solid lines in Figures 10a and 10b). When the derivative is infinite, the plot for $L_{GS}(s)$ approaches zero with an almost vertical slope. The reason is as follows: Since the Jacobi solutions $\mathbf{u}_1$ and $\mathbf{u}_2$ are normal to the normalized ray direction $\mathbf{r}$ (which is a vector of a unit length), geometric spreading can be presented as (equation 26 of Part V),

$$L_{GS}(s) = v_{J,S}\sqrt{v_{\text{ray}}(s)/v_{\text{phs}}(s)|J(s)|} = v_{J,S}\sqrt{v_{\text{ray}}(s)/v_{\text{phs}}(s)}\sqrt{\mathbf{u}_1 \times \mathbf{u}_2 \cdot \mathbf{r}}$$
$$= v_{J,S}\sqrt{v_{\text{ray}}(s)/v_{\text{phs}}(s)}\sqrt{u_1(s)u_2(s)\sin\theta_u(s)} \quad , \quad (12)$$

where $v_{J,S}$ is the so-called conversion velocity, $u_1$ and $u_2$ are the absolute values of vectors $\mathbf{u}_1$ and $\mathbf{u}_2$, respectively, and $\theta_u$ is the angle between these vectors. We assume that the value under the square root is positive. In order to estimate the order of a caustic, we compute the arclength derivative of the geometric spreading,

$$\dot{L}_{GS} = \frac{dL_{GS}}{ds} = \frac{v_{J,S}}{2v_{\text{phs}}(s)} \frac{\dot{v}_{\text{ray}}(s)v_{\text{phs}}(s) - v_{\text{ray}}(s)\dot{v}_{\text{phs}}(s)}{\sqrt{v_{\text{phs}}(s)v_{\text{ray}}(s)}} \cdot \sqrt{u_1(s)u_2(s)\sin\theta_u(s)}$$
$$+ \frac{v_{J,S}}{2}\sqrt{\frac{v_{\text{ray}}(s)}{v_{\text{phs}}(s)}} \frac{\dot{u}_1 u_2 \sin\theta_u + u_1 \dot{u}_2 \sin\theta_u + u_1 u_2 \cos\theta_u \dot{\theta}_u}{\sqrt{u_1 u_2 \sin\theta_u}} \quad . \quad (13)$$

In the proximity of a caustic of any order, the first term including the ray Jacobian, $J = u_1 u_2 \sin\theta_u$, becomes infinitesimal and can be ignored, as the derivative of the geometric spreading, $\dot{L}_{GS}$, includes also finite and/or unbounded terms,

$$\underbrace{\dot{L}_{GS}}_{\text{near caustic}} \approx \frac{v_{J,S}}{2}\sqrt{\frac{v_{\text{ray}}(s)}{v_{\text{phs}}(s)}}\left(\dot{u}_1\sqrt{\frac{u_2 \sin\theta_u}{u_1}} + \dot{u}_2\sqrt{\frac{u_1 \sin\theta_u}{u_2}} + \cos\theta_u \dot{\theta}_u\sqrt{\frac{u_1 u_2}{\sin\theta_u}}\right) \quad . \quad (14)$$

In the proximity of a first-order (line) caustic, either the magnitude $u_1$ becomes small, or $u_2$, or the two vectors become almost collinear (dependent), thus the angle $\theta_u$ becomes small. The derivatives $\dot{u}_1, \dot{u}_2, \dot{\theta}_u$ are normally not small. For each of the three cases mentioned above, the denominator of one of the items in the brackets on the right-hand side of equation 14 becomes infinitesimal, and the whole derivative $\dot{L}_{GS}$ becomes unbounded. Thus, in the proximity of a line caustic, the derivative $\dot{L}_{GS}$ is unbounded.

Now consider the case of a point caustic proximity, where both $u_1$ and $u_2$ are small. In this case, the last item in the brackets in equation 14 does not contribute. Assume that $u_1$ and $u_2$ approach zero with the same rate, i.e. represent infinitesimal values of the same order. In this case their ratio remains finite (not unbounded and not infinitesimal), and the two first bracketed items in equation 14 yield finite contributions to the derivative of the geometric spreading. Thus, in the proximity of a point caustic, the derivative $\dot{L}_{GS}$ is normally bounded. Of course, this is only a tendency, not a rule, because $u_1$ and $u_2$ may be also infinitesimal values of different orders; in this case, either $u_1/u_2$ or $u_2/u_1$ is large, and then $\dot{L}_{GS}$ in the neighborhood of a point caustic becomes unbounded. Still, we suggest the following "rule of thumb": A caustic with an unbounded arclength derivative of geometric spreading is a candidate for the first-order (line) caustic, and that with a finite derivative – for the second-order (point) caustic. Of course, this assumption should be further verified and approved or disproved, and in the case of a line caustic, the line direction can be established.

Example 5: Diving ray in anisotropic ellipsoidal model

Example 5a: Anisotropic ellipsoidal model with constant tilted reference velocity gradient

The model includes an anisotropic ellipsoidal medium (Appendix G) with a background velocity $v$ that changes linearly in space and has a tilted gradient as in Example 1b. The surface midpoint velocity is, $v_a = 3\,\text{km/s}$, the gradient components are, $k_1 = 0.2\,\text{s}^{-1}$, $k_3 = 0.8\,\text{s}^{-1}$, and the offset is, $h = 10\,\text{km}$. The axial "crystal" velocities of the medium at any point are proportional to the background velocity (factorized anisotropic inhomogeneous medium, FAI),

$$A_v(\mathbf{x}) = \lambda_a v(\mathbf{x}) \quad, \quad B_v(\mathbf{x}) = \lambda_b v(\mathbf{x}) \quad, \quad C_v(\mathbf{x}) = \lambda_c v(\mathbf{x}) \quad, \quad v = v_a + \mathbf{k} \cdot \mathbf{x} \quad, \tag{15}$$

where $\lambda_a = 1.25$, $\lambda_b = 1.15$, $\lambda_c = 1$ are the unitless constant values, i.e., the vertical axial velocity coincides with the reference velocity. The two horizontal axial velocities are higher than the vertical velocity. This medium can be viewed as an acoustic orthorhombic medium, whose six parameters are,

$$v_P = \lambda_c v(\mathbf{x}) \quad, \quad \sqrt{1 + 2\varepsilon_2} = \lambda_a \quad, \quad \sqrt{1 + 2\varepsilon_1} = \lambda_b \quad, \quad \eta_1 = \eta_2 = \eta_3 = 0 \quad, \tag{16}$$

where $v_P$ is the vertical (compressional) velocity, $\varepsilon_1$ and $\varepsilon_2$ are the Tsvankin (1997) anisotropy parameters, and $\eta_1, \eta_2, \eta_3$ are the anellipticities (Alkhalifah, 2003; Tsvankin and Grechka, 2011). The ray path is presented in Table 14 and Figure 11, the source-receiver traveltime Hessian – in Table 15, the numerical path characteristics, along with their error values – in Table 16. The accuracy of the numerical traveltime is 11-12 digits. According to the theory explained in Appendix G, the stationary ray path represents a symmetric elliptic arc, with the following horizontal and vertical semi-axes, eccentricity and maximum penetration depth,

$$A_e = 6.853660062 \, \text{km} \,, \qquad B_e = 5.482928050 \, \text{km} \,,$$
$$m_e = 0.6 \,, \qquad\qquad z_{\max} = 1.732928050 \, \text{km} \,. \tag{17}$$

The numerical finite-element solution for the ray path is plotted in Figure 11a. In Figures 11b and 11c, we plot the non-normalized and normalized (divided by $\sigma$) geometric spreading, respectively. Unlike the isotropic medium with a constant gradient considered in Examples 1a and 1b, where the geometric spreading is equal to $\sigma$, this is not so even for the simplest anisotropic case. In order to validate the reciprocity characteristic of the geometrical spreading, the DRT was performed twice: first for $x_{1,S} = -5$ km, $x_{1,R} = +5$ km (solid lines in Figures 11b and 11c), and then the source and receiver were swapped (dashed lines in Figures 11b and 11c). Indeed, at the destination point, the solid and dash lines meet due to the reciprocity of the geometric spreading, but they are different elsewhere through the asymmetric velocity field (wrt the vertical axis of symmetry of the elliptic path). In Table 17, we list the kinematic characteristics, the phase and ray velocities at the source, the conversion velocity $v_{J,S}$, the source-point normalized geometric spreading $(L_{GS}/\sigma)_S$, and the geometric spreading at the destination point, for the forward and reverse paths (when the source and receiver are swapped). The analytical prediction of the latter yields an excellent match with the numerical value. The geometric spreading is very close to an integer number, $L_{GS} = 46 \, \text{km}^2/\text{s}$; most probably, this is an exact theoretical value.

Example 5b: Ellipsoidal model with varying tilted reference velocity gradient

To test the derived relationships for the conversion velocity $v_{J,S}$, and for the source-point value of the normalized geometric spreading, we consider one more example with the ellipsoidal orthorhombic model, with varying gradient of the reference velocity, which leads to asymmetric path. The reference velocity reads,

$$v(\mathbf{x}) = v_a + k_1 x_1 + k_3 x_3 + k_{11} x_1^2 + k_{33} x_3^2 + k_{13} x_1 x_3 \quad , \tag{18}$$

where the components of the spatial gradient and Hessian of the reference velocity are,

$$\begin{aligned} v_a &= 3 \text{ km/s}, & k_{11} &= +0.05 (\text{km} \cdot \text{s})^{-1}, \\ k_1 &= +0.2 \text{ s}^{-1}, & k_{33} &= +0.04 (\text{km} \cdot \text{s})^{-1}, \\ k_3 &= +0.8 \text{ s}^{-1}, & k_{13} &= -0.06 (\text{km} \cdot \text{s})^{-1}. \end{aligned} \tag{19}$$

The origin of the reference frame is located at the midpoint. The computed numerical ray path is shown in Figure 12 and Table 18. The path is asymmetric, with different take-off angles at the source and receiver. The computed endpoint Hessian is presented in Table 19. In Table 20, we list the kinematic and dynamic characteristics of the ray path. As we see, there is an excellent match between the two methods for computing the geometric spreading and for the forward and reverse paths. The ray path is plotted in Figure 12a. The non-normalized and normalized geometric spreading are plotted in Figures 12b and 12c, respectively. The graphs related to the forward path are plotted with solid lines, and those related to the reverse path (after swap of the source and receiver) – with dashed lines.

## CONCLUSIONS

In this last part of our Eigenray study, we present the finite-element solver for the Jacobi DRT equation, needed to compute the geometric spreading along a stationary ray and to identify (and classify) possible caustics. The solver is valid for 3D smooth heterogeneous general anisotropic media and for all types of wave modes (although in this work we only implemented the method for compressional waves). Instead of the traditional initial-value numerical integration approach (e.g., Runge-Kutta) for the DRT, where the derivatives of the unknown function(s) (e.g., paraxial normal shifts at each point along the central ray) are approximated by finite differences, we solve the ODE set with an accurate finite-element implementation using (naturally) the same discretization and the same Hermite interpolation scheme used for the KRT in Part III of this study. This is particularly important for the type of rays studied in this work, involving caustics phenomena along waves traveling in complex geological areas. One of the main advantages of the proposed method is that the resolving matrix of the DRT yields a linear algebraic equation set which coincides with the traveltime Hessian matrix of the stationary ray. This traveltime Hessian matrix has already been computed as part of the solution for the stationary ray path, and hence should not be recomputed, thus making the implementation of the proposed method efficient. We have successfully demonstrated the high accuracy of the proposed method with a number of isotropic and anisotropic benchmark problems.

## ACKNOWLEDGEMENT

The authors are grateful to Emerson for financial and technical support for this study and for permission to publish its results. The gratitude is extended to Ivan Pšenčík, Einar Iversen, Michael Slawinski, Alexey Stovas, Vladimir Grechka, and our colleague Beth Orshalimy, whose valuable remarks helped to improve the content and style of this paper.

## APPENDIX A. FINITE-ELEMENT FORMULATION FOR THE JACOBI DRT SOLVER

The finite-element solver is used to find a single normal solution with arbitrary initial or boundary conditions, which may (or may not) coincide with the initial conditions for any of the four basic solutions of the Jacobi equation. In case they don't coincide, the solution becomes a linear combination of the basic solutions. The specific initial conditions needed for the basic solutions of point-source and plane-wave paraxial rays are discussed in the Appendix D of Part V. Recall that for point-source initial conditions, only the first two basic solutions are used.

In this appendix we derive the weak finite-element formulation in order to solve the second-order Jacobi equation (equation 1). Weak formulation means relaxation in the continuity of the solution, reducing the ODE to the first order, local, weighted residual, linear algebraic equation set. The proposed finite-element discretization is the one used to obtain the Eigenray stationary path, with the Hermite interpolation between the nodes. Thus, both the nodal function values and the nodal derivatives wrt the arclength are independent DoF. The Galerkin method belongs to the class of the weighted-residual methods, where the differential equation is normal to each of the weight functions. We multiply the Jacobi equation 1 by a weight (test) function $w(s)$ and integrate over the element length,

$$\int_{s_{ini}}^{s_{fin}} \frac{d}{ds}\left(L_{\mathbf{rx}} \cdot \mathbf{u} + L_{\mathbf{rr}} \cdot \dot{\mathbf{u}}\right) w(s)\,ds = \int_{s_{ini}}^{s_{fin}} \left(L_{\mathbf{xx}} \cdot \mathbf{u} + L_{\mathbf{xr}} \cdot \dot{\mathbf{u}}\right) w(s)\,ds \qquad , \tag{A1}$$

where $s_{ini}$ and $s_{fin}$ are the values of the arclength at the endpoints of a finite element, $s_{ini} < s_{fin}$. The operator in equation A1 (the weak formulation) is applied locally within a single finite element, i.e., a region where the Hermite interpolation is carried out by continuous polynomials. Within a single finite element, the arclength is mapped onto the internal flow parameters, $\xi$, $-1 \leq \xi \leq +1$,

$$s = s(\xi), \quad s(-1) = s_{\text{ini}}, \quad s(+1) = s_{\text{fin}} \quad , \tag{A2}$$

and a positive metric $s' = ds/d\xi$ exists. Note that the arclength is only used in the derivation stage, while the final integration formulae are all in terms of the internal flow parameter $\xi$.

Consider the left side of equation A1 and apply integration by parts,

$$\begin{aligned}
\int_{s_{\text{ini}}}^{s_{\text{fin}}} \frac{d}{ds}\left(L_{\mathbf{rx}} \cdot \mathbf{u} + L_{\mathbf{rr}} \cdot \dot{\mathbf{u}}\right) w(s)\, ds &= \int_{s_{\text{ini}}}^{s_{\text{fin}}} w(s)\, d\left(L_{\mathbf{rx}} \cdot \mathbf{u} + L_{\mathbf{rr}} \cdot \dot{\mathbf{u}}\right) = \\
&= \left(L_{\mathbf{rx}} \cdot \mathbf{u} + L_{\mathbf{rr}} \cdot \dot{\mathbf{u}}\right) w(s)\Big|_{s_{\text{ini}}}^{s_{\text{fin}}} - \int_{s_{\text{ini}}}^{s_{\text{fin}}} \left(L_{\mathbf{rx}} \cdot \mathbf{u} + L_{\mathbf{rr}} \cdot \dot{\mathbf{u}}\right) dw = \\
&= \left(L_{\mathbf{rx}} \cdot \mathbf{u} + L_{\mathbf{rr}} \cdot \dot{\mathbf{u}}\right) w(s)\Big|_{s_{\text{ini}}}^{s_{\text{fin}}} - \int_{s_{\text{ini}}}^{s_{\text{fin}}} \left(L_{\mathbf{rx}} \cdot \mathbf{u} + L_{\mathbf{rr}} \cdot \dot{\mathbf{u}}\right) \dot{w}(s)\, ds \quad ,
\end{aligned} \tag{A3}$$

where

$$\dot{w}(s) = \frac{dw(s)}{ds} = \frac{dw(\xi)}{d\xi} \frac{d\xi}{ds} = \frac{dw(\xi)/d\xi}{ds(\xi)/d\xi} = \frac{w'(\xi)}{s'(\xi)} \quad . \tag{A4}$$

We recall that the dot symbol means a derivative wrt the arclength, while prime means a derivative wrt the internal parameter $\xi$. Combining equations A1 – A4, we obtain,

$$\int_{\xi=-1}^{\xi=+1} \left(L_{\mathbf{rx}} \cdot \mathbf{u} + L_{\mathbf{rr}} \cdot \dot{\mathbf{u}}\right) w'\, d\xi + \int_{\xi=-1}^{\xi=+1} \left(L_{\mathbf{xx}} \cdot \mathbf{u} + L_{\mathbf{xr}} \cdot \dot{\mathbf{u}}\right) w s'\, d\xi = \left(L_{\mathbf{rx}} \cdot \mathbf{u} + L_{\mathbf{rr}} \cdot \dot{\mathbf{u}}\right) w \Big|_{\xi=-1}^{\xi=+1} \quad . \tag{A5}$$

We note that the weak formulation in equation A5 does not include the second derivative $\ddot{\mathbf{u}}$. The nodal values of the solution and its derivative are unknown, and the Hermite interpolation is applied between the nodes,

$$\mathbf{u}(\xi) = \sum_{I=1}^{n} \mathbf{u}_I h_I(\xi) + \sum_{I=1}^{n} \mathbf{u}'_I d_I(\xi) \ ,$$

$$\mathbf{u}'(\xi) = \sum_{I=1}^{n} \mathbf{u}_I h'_I(\xi) + \sum_{I=1}^{n} \mathbf{u}'_I d'_I(\xi) \ ,$$

(A6)

where $n = 2,3$ for the two-node or three-node elements, respectively. The shape functions $h(\xi)$ and $d(\xi)$ are listed in equation sets A3 and A6 of Part III for two-node and three-node elements, respectively. The Galerkin method is also applied locally within each finite element, and yields a local, first-order, weighted residual, linear algebraic equation set.

Similarly to equation A4, the nodal derivatives wrt the arclength may be rescaled to the corresponding derivatives wrt the internal parameter,

$$\mathbf{u}(\xi) = \sum_{I=1}^{n} \mathbf{u}_I h_I(\xi) + \sum_{I=1}^{n} \dot{\mathbf{u}}_I \ s'_I d_I(\xi) \ ,$$

$$\dot{\mathbf{u}}(\xi) = \sum_{I=1}^{n} \mathbf{u}_I \frac{h'_I(\xi)}{s'(\xi)} + \sum_{I=1}^{n} \dot{\mathbf{u}}_I \frac{s'_I d'_I}{s'(\xi)}(\xi) \ .$$

(A7)

Function $s'(\xi) = \sqrt{\mathbf{x}'(\xi) \cdot \mathbf{x}'(\xi)}$ (the derivative of the arclength $s$ wrt the internal parameter $\xi$) is the metric of the finite element, which is known at the nodal and non-nodal points; $s'_I$ are the nodal values of the metric. With the Galerkin method, widely used in the finite-element approach, the test (weight) functions $w(\xi)$ are the same as the Hermite interpolation functions $h(\xi)$ and $d(\xi)$ presented in Appendix A of Part III. For example, for a three-node element with end nodes $A, C$ and a central node $B$, the weight $w(\xi)$ runs all possible interpolation functions successively in the following sequence,

$$w(\xi) = \{h_a(\xi), \ d_a(\xi), \ h_b(\xi), \ d_b(\xi), \ h_c(\xi), \ d_c(\xi)\} \quad . \tag{A8}$$

Thus, there are a total number of $2n$ interpolation functions. Introducing each test function into equation A5, we obtain a vector-form equation set of dimension $6n$. It can be arranged in matrix form, because the nodal values of the solution and their derivatives can be moved outside the integral sign,

$$\mathbf{S}_{\text{loc}} \hat{\mathbf{u}}_{\text{loc}} = \mathbf{f}_{\text{loc}} \quad , \tag{A9}$$

where vector $\hat{\mathbf{u}}$ includes both $\mathbf{u}$ and $\dot{\mathbf{u}}$. For a three-node element, $\hat{\mathbf{u}}$ is a vector of length 18,

$$\hat{\mathbf{u}} = \begin{bmatrix} \mathbf{u}_a & \dot{\mathbf{u}}_a & \mathbf{u}_b & \dot{\mathbf{u}}_b & \mathbf{u}_c & \dot{\mathbf{u}}_c \end{bmatrix} \quad , \tag{A10}$$

and each block on the right side is a vector of length 3. The subscript "loc" in equation A9 emphasizes that a single element is under consideration. The local "stiffness" matrix $\mathbf{S}_{\text{loc}}$ (in analogy with the finite element analysis of elastic mechanical structures) consists of $n \times n$ blocks, each of dimension $6 \times 6$. Each block of the matrix is related to nodes $I$ and $J$. The local (normal) "displacement" $\mathbf{u}_{\text{loc}}$ and the local "load" $\mathbf{f}_{\text{loc}}$ are vectors of $n$ blocks, where the length of each block is 6. The displacements are unknown, while the stiffness and load can be computed from the ray path. We recall that $n$ is the number of nodes in a single finite element.

Introduction of equations A9 and A10 into the left-hand side of equation A5 yields the blocks of the local stiffness matrix. A single $6 \times 6$ block consists of four sub-blocks of dimension $3 \times 3$,

$$\mathbf{S}_{IJ} = \begin{bmatrix} \mathbf{S}_{\mathbf{x}_I \mathbf{x}_J} & \mathbf{S}_{\mathbf{x}_I \mathbf{r}_J} \\ \mathbf{S}_{\mathbf{r}_I \mathbf{x}_J} & \mathbf{S}_{\mathbf{r}_I \mathbf{r}_J} \end{bmatrix} \quad . \tag{A11}$$

For the sake of symmetry, we multiply the second row of the sub-blocks in equation A11 by $s'_I$. This multiplication does not affect the right-hand vector in equation A9, because the corresponding components of this vector are zero. The sub-blocks of the local stiffness block become,

$$\mathbf{S}_{\mathbf{x}_I \mathbf{x}_J} = \int_{-1}^{+1} \left( L_{\mathbf{xx}} s' h_I h_J + L_{\mathbf{xr}} h_I h'_J + L_{\mathbf{rx}} h'_I h_J + \frac{L_{\mathbf{rr}}}{s'} h'_I h'_J \right) d\xi \quad ,$$

$$\mathbf{S}_{\mathbf{x}_I \mathbf{r}_J} = \int_{-1}^{+1} \left( L_{\mathbf{xx}} s' h_I d_J + L_{\mathbf{xr}} h_I d'_J + L_{\mathbf{rx}} h'_I d_J + \frac{L_{\mathbf{rr}}}{s'} h'_I d'_J \right) d\xi \, s'_J \quad ,$$

$$\mathbf{S}_{\mathbf{r}_I \mathbf{x}_J} = \int_{-1}^{+1} \left( L_{\mathbf{xx}} s' d_I h_J + L_{\mathbf{xr}} d_I h'_J + L_{\mathbf{rx}} d'_I h_J + \frac{L_{\mathbf{rr}}}{s'} d'_I h'_J \right) d\xi \, s'_I \quad ,$$

$$\mathbf{S}_{\mathbf{r}_I \mathbf{r}_J} = \int_{-1}^{+1} \left( L_{\mathbf{xx}} s' d_I d_J + L_{\mathbf{xr}} d_I d'_J + L_{\mathbf{rx}} d'_I d_J + \frac{L_{\mathbf{rr}}}{s'} d'_I d'_J \right) d\xi \, s'_I s'_J \quad .$$

(A12)

Recall that the matrices $L_{\mathbf{xx}}(\xi), L_{\mathbf{xr}}(\xi), L_{\mathbf{rx}}(\xi), L_{\mathbf{rr}}(\xi)$ and the metric $s'(\xi)$ are known for any $\xi$ from the stationary ray path solution. The diagonal stiffness blocks of the local stiffness matrix are symmetric, while the off-diagonal blocks are transposed to their counterparts, $\mathbf{S}_{IJ} = \mathbf{S}_{JI}^T$ (for both $I = J$ and $I \neq J$). Hence, the entire local and global stiffness matrices are symmetric.

Introduction of equations A9 and A11 into the right-hand side of equation A5 yields the blocks of the local load vector. However, only two weight functions contribute to the load. There is a single nonzero weight function at the left end of the element, $\xi = -1$, and a single nonzero weight function at the right end, $\xi = +1$. Both functions accept value 1 at their ends. Each load block of length 6 consists of two sub-blocks of length 3, $\mathbf{f}_\mathbf{x}$ and $\mathbf{f}_\mathbf{r}$. The first sub-block is

Page 27 of 100

related to the Jacobi set solution, and the second to its derivative. One can call them "forces" and "moments", respectively, or generally "loads", in analogy with mechanical structure problems. The moment sub-blocks of all the blocks are zero because the interpolation functions $d_I(\xi)$ vanish at both endpoints, $\mathbf{f_r} = 0$.

The "load" of a two-node element consists of two blocks of length 6, and the "load" of a three-node element consists of three blocks of length 6. Following the statement above, the central load block of a three-node element is zero. The two other blocks look alike, no matter whether the element is two-node or three-node. For a three-node element, the force-based sub-blocks of the element end nodes read,

$$\mathbf{f}_{\mathbf{x},a} = \underbrace{-L_{\mathbf{rx}}\mathbf{u}_a - L_{\mathbf{rr}}\dot{\mathbf{u}}_a}_{\xi=-1} \quad , \quad \mathbf{f}_{\mathbf{x},c} = \underbrace{+L_{\mathbf{rx}}\mathbf{u}_c + L_{\mathbf{rr}}\dot{\mathbf{u}}_c}_{\xi=+1} \quad . \tag{A13}$$

For a two-node element, we just replace index $c$ by $b$.

Finally, we add (stack) equations of type A9 for all elements, leading to,

$$\mathbf{S}_{\text{glb}}\hat{\mathbf{u}}_{\text{glb}} = \mathbf{f}_{\text{glb}} \quad , \tag{A14}$$

where the subscript "glb" means global, and $\hat{\mathbf{u}}_{\text{glb}}$ includes all nodal components of the Jacobi DRT set solution and the derivatives of these components wrt the arclength. At the joints, the Jacobi DRT solution components, function $\mathbf{u}$ and its derivative $\dot{\mathbf{u}}$, are continuous, and the corresponding overlapping parts of the global stiffness matrix and global load vector are added at the joints. For the load vector, this means that $\mathbf{f}_{\mathbf{x},c}$ of the previous element is added to $\mathbf{f}_{\mathbf{x},a}$ of the next element; the result is zero, as they compensate each other. Eventually, only $\mathbf{f}_{\mathbf{x},a}$ of the first element and $\mathbf{f}_{\mathbf{x},c}$ of the last element of the whole structure contribute to the global load vector, and equation A13 can be arranged as,

$$\mathbf{f}_{\mathbf{x}}^S = -L_{\mathbf{rx}}^S \mathbf{u}_S - L_{\mathbf{rr}}^S \dot{\mathbf{u}}_S \quad , \quad \mathbf{f}_{\mathbf{x}}^R = +L_{\mathbf{rx}}^R \mathbf{u}_R + L_{\mathbf{rr}}^R \dot{\mathbf{u}}_R \quad , \tag{A15}$$

where indices $S$ and $R$ are related to the source and receiver, respectively. Due to the boundary or initial conditions, we know two (and only two) of the four endpoint values $\mathbf{u}_S, \dot{\mathbf{u}}_S, \mathbf{u}_R, \dot{\mathbf{u}}_R$; the other two are unknown.

Introducing equation A2 for $L_{\mathbf{xx}}, L_{\mathbf{xr}}, L_{\mathbf{rx}}, L_{\mathbf{rr}}$ into equation A12 for the block of the local stiffness matrix $\mathbf{S}_{IJ}$, and taking into account that $l(\xi) \equiv 2\sqrt{\mathbf{x}'(\xi)\mathbf{x}'(\xi)} = 2s'(\xi)$ (recall that prime means a derivative wrt the internal flow parameter $\xi$), we conclude that the stiffness block of dimension $6 \times 6$ completely coincides with the corresponding block of the traveltime Hessian derived for the kinematics (equations F19 – F22 of Part III). This in turn, means that the whole local stiffness matrix $12 \times 12$ or $18 \times 18$ (depending on the element type) coincides with the local traveltime Hessian. After assembly of the local matrices, the global stiffness coincides with the global Hessian, provided the BC of the Eigenray (i.e., the source and receiver locations) have not yet been implemented (because the Jacobi DRT set has its own IC). The boundary conditions of the kinematic Eigenray formulation modify the Hessian of the first and last finite elements. This change does not match the IC of the dynamic Eigenray formulation. Thus, the global stiffness matrix reads,

$$\mathbf{S}_{\text{glb}} = \nabla_\mathbf{d} \nabla_\mathbf{d} t \quad , \tag{A16}$$

where subscript **d** means all DoF. This means that the global traveltime Hessian $\nabla_\mathbf{d}\nabla_\mathbf{d} t$ obtained in the last iteration of the Eigenray optimization (corresponding to the stationary path) should be used as the global stiffness matrix of the Jacobi DRT equation. This is a remarkable result which makes our Eigenray method very attractive – the traveltime Hessian of the stationary ray is the "stiffness" matrix to be used for the dynamic solution.

Next, we can transfer the right-hand term of equation A14, whose source and receiver components are listed in equation A15 (and whose other components vanish) to the left side of equation A14, and the coefficients in equation A15 yield additional contributions to the corresponding components of the global stiffness matrix $\mathbf{S}_{\text{glb}}$,

$$\begin{aligned}
\mathbf{S}_{\text{glb}}(0:2, 0:2) &+= L^S_{\mathbf{rx}} \ , \\
\mathbf{S}_{\text{glb}}(0:2, 3:5) &+= L^S_{\mathbf{rr}} \ , \\
\mathbf{S}_{\text{glb}}(6N:6N+2, 6N:6N+2) &-= L^R_{\mathbf{rx}} \ , \\
\mathbf{S}_{\text{glb}}(6N:6N+2, 6N+3:6N+5) &-= L^R_{\mathbf{rr}} \ ,
\end{aligned} \quad (A17)$$

where $N+1$ is the total number of nodes (enumerated from zero to $N$). In this equation set, notations $\mathbf{S}_{\text{glb}}(m_1:m_2, n_1:n_2)$ mean a block of the global stiffness matrix whose rows run from $m_1$ through $m_2$ and columns from $n_1$ through $n_2$. All blocks in this equation set are $3\times 3$. Symbols $'+='$ and $'-='$ mean "add to" or "subtract from" (respectively) the corresponding blocks of the global stiffness matrix that was obtained after assembly, but before transferring the load vector to the left side of equation A14. The indices of the rows and columns in the stiffness matrix start at zero. The first two equations of set A17 are related to the front end of the finite-element structure, and the last two equations to its rear end. After transferring the boundary conditions to the left side, the left-hand side of the global equilibrium equation A13 becomes $\hat{\mathbf{S}}_{\text{glb}} \cdot \hat{\mathbf{u}}_{\text{glb}}$, while the right-hand side is zero. We apply notation $\hat{\mathbf{S}}_{\text{glb}}$ for the global stiffness matrix subject to the operator of equation A17,

$$\hat{\mathbf{S}}_{\text{glb}} \cdot \hat{\mathbf{u}}_{\text{glb}} = 0 \qquad . \tag{A18}$$

Note that this operator (equation A17) ruins the symmetry of the stiffness matrix $\mathbf{S}_{\text{glb}}$, and matrix $\hat{\mathbf{S}}_{\text{glb}}$ is no longer identical to the global traveltime Hessian $\nabla_{\mathbf{d}} \nabla_{\mathbf{d}} t$.

As mentioned, the Jacobi DRT equation set is not fully independent, and even after implementing the boundary conditions (replacing the known values of the shifts and their arclength derivatives by numbers), the determinant of the global stiffness vanishes, due to an indefinite and meaningless tangential Jacobi solution. To obtain a determined equation system, we augment two equations of set 4 per each node to equation set A18. The second equation of set 4 includes the curvature vector $\dot{\mathbf{r}}$,

$$\begin{aligned}\dot{\mathbf{r}} &= \frac{d\mathbf{r}}{ds} = \frac{d}{ds} \frac{\mathbf{x}'}{\sqrt{\mathbf{x}' \cdot \mathbf{x}'}} = \frac{1}{\sqrt{\mathbf{x}' \cdot \mathbf{x}'}} \frac{d}{d\xi} \frac{\mathbf{x}'}{\sqrt{\mathbf{x}' \cdot \mathbf{x}'}} \\ &= \frac{\mathbf{x}''}{\mathbf{x}' \cdot \mathbf{x}'} - \frac{\mathbf{x}'(\mathbf{x}' \cdot \mathbf{x}'')}{(\mathbf{x}' \cdot \mathbf{x}')^2} = \frac{\mathbf{I}(\mathbf{x}' \cdot \mathbf{x}') - \mathbf{x}' \otimes \mathbf{x}'}{(\mathbf{x}' \cdot \mathbf{x}')^2} \mathbf{x}'' \qquad .\end{aligned} \tag{A19}$$

This vector can be computed at the nodes of all elements, applying the Eigenray solution for the stationary path and equation E8 of Part III. We note that the ray path solutions have been obtained in Part III with $C_1$ continuity, which means that for smooth media, the path locations $\mathbf{x}$ and directions $\mathbf{r}$ are continuous at the joints, but the curvatures $\dot{\mathbf{r}}$ are not. (Recall that in Part III we also discuss the options to impose discontinuity in some of the nodes, like interface nodes, but this is beyond the scope of this study). Therefore, for the continuity conditions, we take the average curvature $\dot{\mathbf{r}}$ for two adjacent elements at the joints and introduce it into equation 4. We further note that vector $\mathbf{r}$ is normalized to the unit length, which means that it can change only its direction, $\mathbf{r} \cdot \dot{\mathbf{r}} = 0$. This in turn, means that only the magnitude of the curvature can be discontinuous at the joints, while its direction is continuous. The "jumps" in the curvature magnitude decay as the number of finite elements increases and the numerical Eigenray solution converges to the exact path.

Thus, to solve the Jacobi equation set, we minimize the quadratic target function that follows from equation A18,

$$T = \hat{\mathbf{u}}_{\text{glb}} \cdot \hat{\mathbf{S}}_{\text{glb}}^T \hat{\mathbf{S}}_{\text{glb}} \cdot \hat{\mathbf{u}}_{\text{glb}} \qquad , \qquad (A20)$$

where the stiffness matrix $\hat{\mathbf{S}}_{\text{glb}}$ includes additions/subtractions of equation set A17. The target function in equation A20 is subjected to the initial conditions and to additional constraints of equation 4, enforcing the normal solution (two constraints per node). Constraints related to the DoF of the IC may be omitted: We assume that the IC are normal to the ray (see the next appendix for the initial conditions). Since the target function is quadratic, this constrained minimization is equivalent to an augmented linear set (which includes the constraints), with a unique solution. Theoretically, the target function $T$ should approach zero, but in practice it has a fairly small positive value, due to the above-mentioned discontinuity of the ray path curvature at the joints, and other numerical inaccuracies.

## APPENDIX B. DIVING RAY FOR CONSTANT VELOCITY GRADIENT

Vertical velocity gradient of isotropic model

A linear velocity model is specified by the surface velocity $v_a$ and the constant vertical velocity gradient $k$. The ray trajectory is a circular arc, and the traveltime reads (Červený, 2000),

$$k t = \operatorname{arccosh}\left[1 + \frac{h^2 + (\tilde{z}_R - \tilde{z}_S)^2}{2\tilde{z}_S \tilde{z}_R}\right] \quad , \tag{B1}$$

where $k$ is the vertical velocity gradient, $h$ is the source-receiver horizontal offset, and $\tilde{z}_S, \tilde{z}_R$ are absolute depths of the source and receiver, measured from the zero-velocity level (above the earth's surface). Assuming the surface velocity is $v_a$, these levels for a diving ray are,

$$\tilde{z}_S = \tilde{z}_R = v_a / k \quad , \tag{B2}$$

and the traveltime equation simplifies to,

$$k t = \operatorname{arccosh}\left(1 + \frac{k^2 h^2}{2 v_a^2}\right) \quad . \tag{B3}$$

The radius of trajectory, maximum penetration depth and normalized arclength reads,

$$\rho = \frac{\sqrt{4 v_a^2 + k^2 h^2}}{2k} \quad , \quad z_{\max} = \rho - \frac{v_a}{k} \quad , \quad \frac{s}{\rho} = \theta_R - \theta_S = \pi - 2\theta_S \quad . \tag{B4}$$

The take-off angles $\theta_S$ and $\theta_R = \pi - \theta_S$ at the source and receiver points read,

$$\sin \theta_S = \sin \theta_R = v_a p_h = v_a \frac{dt}{dh} = \frac{v_a}{k\rho} \quad , \tag{B5}$$

where $p_h$ is the invariant horizontal slowness. Combining equations B3 and B4 with equations 2 – 4 of Part IV, we obtain the geometric spreading of the diving ray,

$$L_{GS} = k\rho h \quad . \tag{B6}$$

Note that for the constant gradient velocity model, parameter $\sigma$ defined in equation 8 is equal to the geometric spreading $L_{GS}$.

Tilted velocity gradient of isotropic model

Next we consider a diving ray in a model with a tilted velocity gradient, with constant Cartesian components $k_1$ and $k_3$, where the lateral component $k_1$ may be of any sign and the vertical component $k_3$ is assumed positive,

$$v = v_a + k_1 x_1 + k_3 x_3 \tag{B7}$$

The origin of the reference frame is at the midpoint on the surface, and $v_a$ is the velocity at the midpoint. The normalized traveltime reads,

$$k\,t = \ln \frac{\cos\left(\frac{\alpha_R}{2} - \frac{\pi}{4}\right)\cos\left(\frac{\alpha_S}{2} + \frac{\pi}{4}\right)}{\cos\left(\frac{\alpha_S}{2} - \frac{\pi}{4}\right)\cos\left(\frac{\alpha_R}{2} + \frac{\pi}{4}\right)} \tag{B8}$$

where $k$ is the gradient absolute value, $k = \sqrt{k_1^2 + k_3^2}$. Parameters $\alpha_S$ and $\alpha_R$ are signed central angles of the source and receiver on the circular arc, measured from the tilted gradient direction,

$$\alpha_S = \arctan \frac{hk_3^2 + 2v_a k_1}{hk_1 k_3 - 2v_a k_3} \quad , \quad \alpha_R = \arctan \frac{hk_3^2 - 2v_a k_1}{hk_1 k_3 + 2v_a k_3} \tag{B9}$$

The formulae for the radius $\rho$ of the arc, the maximum penetration depth $z_{\max}$ and geometric spreading $L_{GS}$ prove to be the same as in equations B4 and B6, but $k_3$ should be introduced instead of $k$. The arclength reads,

$$s = \rho(\alpha_R - \alpha_S) \quad . \tag{B10}$$

Despite the lateral gradient component, the ray trajectory is symmetric about the vertical line of the midpoint. The take-off angles are,

$$\theta_S = \arctan\frac{2v_a}{k_3 h} \quad , \quad \theta_R = \pi - \theta_S \quad . \tag{B11}$$

The horizontal slowness at the departure and arrival points reads,

$$p_{h,S} = \frac{v_a}{v_S} \cdot \frac{1}{k_3 \rho} \quad , \quad v_S = v_a - \frac{k_1 h}{2} \quad ,$$
$$p_{h,R} = \frac{v_a}{v_R} \cdot \frac{1}{k_3 \rho} \quad , \quad v_R = v_a + \frac{k_1 h}{2} \quad , \tag{B12}$$

where $v_S$ and $v_R$ are the medium velocities at the endpoints. Note that equation 2 of Part IV cannot be applied due to lateral variations in velocity, but $L_{GS} = \sigma$ for any constant gradient model, including that with a tilted gradient.

### APPENDIX C. DIVING RAY FOR CONIC VELOCITY MODEL

With a constant vertical velocity gradient, the medium velocity at infinite depth becomes unbounded, which is unrealistic. Compacted sediments can be adequately described by asymptotically bounded velocity models, such as exponential velocity model (Ravve and Koren, 2006a, 2006b), conic model (Ravve and Koren, 2007), hyperbolic model (Muscat, 1937; Ravve and Koren, 2013) and exponential slowness model (Al-Chalabi, 1997; Robein, 2003). They are described by three parameters: surface velocity $v_a$, surface gradient $k_a$ and asymptotic velocity $v_\infty$. The name "conic" comes from the shape (topology) of ray trajectories which are cross-sections of a conic surface. For the conic velocity model,

$$v(\tilde{z}) = \frac{R\tilde{z}}{\sqrt{1+Q^2\tilde{z}^2}} \quad, \quad \text{where} \quad \tilde{z} = z + h \quad , \tag{C1}$$

and $\tilde{z}$ is the absolute depth below the vanishing velocity level (the origin located above the earth's surface). $R, Q$ and $H$ are internal parameters related to the physical parameters of the model,

$$R = \frac{k_a v_\infty^3}{\left(v_\infty^2 - v_a^2\right)^{3/2}} \quad, \quad Q = \frac{R}{v_\infty} \quad, \quad H = \frac{v_a}{\sqrt[3]{k_a R^2}} \quad . \tag{C2}$$

$R$ has units of velocity gradient, $H$ and $Q$ have units of distance and reciprocal distance, respectively. At the surface, $\tilde{z} = H$. In the conic model, the ray path of a diving ray is an elliptic arc of offset-dependent eccentricity $m_e$,

$$m_e^2(h) = 1 - \frac{2\cos^2\theta_c}{1+\sqrt{1+Q^2 h^2 \cos^4\theta_c}} \quad, \quad 0 \leq m_e < 1 \quad , \tag{C3}$$

where $\theta_c$ is the critical angle,

$$\sin\theta_c = \frac{v_a}{v_\infty} \quad . \tag{C4}$$

The pre-critical rays of the conic model are hyperbolic with straight asymptotes at infinite depth, the critical rays are parabolic with unbounded depth and no asymptote, and the post-critical rays are elliptic – they reach maximum depth and return to the surface. The relationships in this appendix are valid for elliptic rays only. Note that for unbounded asymptotic velocity $v_\infty$, the eccentricity $m_e$ vanishes, and an elliptic trajectory becomes circular: A linear velocity model is a limit case of a conic model.

The horizontal slowness is constant for the whole ray path (this is true for any 1D model),

$$p_h = \frac{1}{m_e v_\infty} \quad . \tag{C5}$$

The take-off angles at the endpoints of the ray path are $\theta_S \equiv \theta_a$ and $\theta_R = \pi - \theta_a$, where

$$\sin \theta_a = \frac{\sin \theta_c}{m_e} \quad , \quad \theta_a > \theta_c \quad . \tag{C6}$$

The major and minor semi-axes $A_e$, $B_e$ of the ellipse, respectively, are given by,

$$A_e^2 = \frac{h^2}{4} + \frac{H^2}{1 - m_e^2} \quad , \quad B_e^2 = A_e^2 \left(1 - m_e^2\right) \quad . \tag{C7}$$

The maximum penetration depth is,

$$z_{\max} = \frac{v_a \cos^2 \theta_c}{k_a} \left( \frac{\cos \theta_c}{\sqrt{\sin^2 \theta_a - \sin^2 \theta_c}} - 1 \right) \quad . \tag{C8}$$

The normalized arclength requires an elliptic integral of the second kind,

$$p_h R s = \frac{m_e^2}{1 - m_e^2} \frac{\sin 2\theta_a}{\sqrt{1 - m_e^2 \sin^2 \theta_a}} + 2 \int_{\theta_a}^{\pi/2} \sqrt{1 - m_e^2 \sin^2 \theta} \, d\theta \quad , \tag{C9}$$

and the normalized traveltime reads,

$$R t = \frac{2 m_e^2}{1 - m_e^2} \frac{\cos \theta_a}{\cos \theta_c} + 2 \operatorname{arctanh} \frac{\cos \theta_a}{\cos \theta_c} \quad . \tag{C10}$$

Introduction of equations C3 and C6 into C10 yields the traveltime as a function of a single variable – offset $h$, and the three constant model parameters: $v_a, k_a$ and $v_\infty$. We compute the first and second derivatives of this function to establish the theoretical geometric spreading.

Note that for any 1D isotropic model, the curvature of the ray trajectory reads (Kaufman, 1953; Ravve and Koren, 2006a, 2006b),

$$\kappa = \frac{d\theta}{ds} = \frac{1}{p_h k(\theta)} \quad , \tag{C11}$$

where the velocity gradient $k$ is a function of depth, but may be considered a function of the ray angle $\theta$ as well. In particular, for the conic velocity model,

$$k(\theta) = R\left(1 - m_e^2 \sin^2\theta\right)^{3/2} \quad , \tag{C12}$$

the horizontal propagation (offset) reads,

$$h = \int_A^B \sin\theta \, ds = \frac{1}{p_h} \int_{\theta_a}^{\theta_b} \frac{\sin\theta \, d\theta}{k(\theta)} \quad . \tag{C13}$$

Parameter $\sigma$, defined in equation 5 of Part V, reads,

$$\sigma = \int_A^B v \, ds = \int_{\theta_a}^{\theta_b} \frac{\sin\theta}{p_h} \frac{d\theta}{p_h k} = \frac{1}{p_h^2} \int_{\theta_a}^{\theta_b} \frac{\sin\theta \, d\theta}{k(\theta)} \quad . \tag{C14}$$

Combining equations C13 and C14, we conclude that for any 1D isotropic model,

$$\sigma = h / p_h \quad . \tag{C15}$$

This relationship is valid for any ray, except strictly vertical. For a vertical ray,

$$\sigma = v_2^2 \Delta t = v_2^2 \Delta z / v_1 \quad , \tag{C16}$$

where $\Delta z$ is the vertical propagation, $v_1 = \Delta z / \Delta t$ is the interval velocity (local average), $v_2$ is the local RMS velocity, and $\Delta t$ is the one-way vertical time.

# APPENDIX D. DIVING RAY IN SIMPLE CAUSTIC-GENERATING VELOCITY MODEL

In this appendix, we study a model that consists of a horizontal layer with a constant velocity $v_a$ and thickness $z_h$, and a half space with constant vertical velocity gradient $k$. At the interface, the velocity is continuous. This model allows caustics, and our goal is to study the caustic criterion.

The ray trajectory has a vertical symmetry line and consists of two straight intervals (in the constant velocity layer) with a circular arc in between (in the half-space with the constant vertical velocity gradient), as shown in Figure 13. Introduce the following notation,

$$\Delta z_v = v_a / k \qquad . \qquad (D1)$$

Let $\theta_a$ be the take-off angle. Then, according to Figure 13, the horizontal offset reads,

$$h = 2z_h \tan \theta_a + 2\Delta z_v \cot \theta_a \qquad . \qquad (D2)$$

Consider a paraxial ray with a take-off angle slightly exceeding that of the central ray. If its offset also exceeds the central ray offset, then the paraxial ray path unavoidably intersects the central ray path somewhere, and a caustic occurs. Thus, the sufficient condition for the caustic is,

$$dh / d\theta_a > 0 \qquad . \qquad (D3)$$

Otherwise, the caustic does not exist, i.e., for this simple model it is also a necessary condition. The caustic criterion becomes,

$$\frac{dh}{d\theta_a} = \frac{2z_h}{\cos^2\theta_a} - \frac{2\Delta z_v}{\sin^2\theta_a} > 0 \quad , \tag{D4}$$

or,

$$\theta_a > \theta_c \quad \text{where} \quad \theta_c = \arctan\sqrt{\Delta z_v / z_h} \quad . \tag{D5}$$

Thus, a critical angle $\theta_c$ exists, such that a caustic appears for any take-off angle exceeding the critical value. At $\theta_a = \theta_c$, the derivative $dh/d\theta_a$ changes its sign from minus to plus; thus, this is the minimum point for $h(\theta_a)$. The minimum offset is,

$$h_{\min} = 2z_h \tan\theta_c + 2\Delta z_v \cot\theta_c = 4\sqrt{z_h \Delta z_v} \quad . \tag{D6}$$

We can normalize the offset,

$$\frac{h}{h_{\min}} = \frac{\sqrt{z_h}}{2\sqrt{\Delta z_v}} \tan\theta_a + \frac{\sqrt{\Delta z_v}}{2\sqrt{z_h}} \cot\theta_a \quad . \tag{D7}$$

For $z_h / \Delta z_v = 3$ (this case is shown in Figure 13), the critical angle is $\theta_c = \pi/6 = 30^\circ$. The normalized offset is plotted in Figure 14a. Thus, for an offset below the minimum value, there are no diving rays. The principle of stationary traveltime gives a solution for any two endpoints, but in this case there is a trivial solution – a straight line on the earth's surface connecting the source and receiver. This straight line is always a local minimum, for any offset. In addition, for $h = h_{\min}$, a single diving ray exists. It follows from equation D7 and Figure 14a that for $h > h_{\min}$, two diving rays co-exist (in addition to the trivial straight path), one of them with no caustics, $\theta_a < \theta_c$, and another with a caustic, for $\theta_a > \theta_c$. To find the take-off angles of both diving rays, we solve equation D7 for the given offset $h$ and unknown $\theta_a$,

$$\frac{h}{h_{\min}} = \frac{\sqrt{z_h}}{2\sqrt{\Delta z_v}} \tan\theta_a + \frac{\sqrt{\Delta z_v}}{2\sqrt{z_h}} \cot\theta_a \qquad . \tag{D8}$$

This leads to a quadratic equation with the roots,

$$\theta_a = \arctan \frac{\sqrt{\Delta z_v}}{\sqrt{z_h}} \frac{h \pm \sqrt{h^2 - h_{\min}^2}}{h_{\min}} \qquad . \tag{D9}$$

The traveltime of the circular arc is given by (Červený, 2000) (Section 3.7.2, paragraph 4),

$$t_{\text{circ}} = \frac{1}{k} \operatorname{arccosh}\left(1 + \frac{k^2 \hat{h}^2}{2 v_a^2}\right) \qquad , \tag{D10}$$

where $\hat{h}$ is the portion of the offset corresponding to the circular part of the ray path,

$$\hat{h} = 2\Delta z_v \cot\theta_a \qquad . \tag{D11}$$

The radius of the circular path,

$$\rho = \frac{\Delta z_v}{\sin\theta_a} \qquad . \tag{D12}$$

the full arclength reads,

$$s = 2\rho(\pi/2 - \theta_a) + \frac{2 z_h}{\cos\theta_a} = \frac{\Delta z_v (\pi - 2\theta_a)}{\sin\theta_a} + \frac{2 z_h}{\cos\theta_a} \qquad . \tag{D13}$$

For each of the two straight intervals,

$$t_{\text{lin}} = \frac{z_h}{v_a \cos \theta_a} \quad . \tag{D14}$$

For the circular arc, the depth $\tilde{z}$ of the endpoints measured relative to the vertical level of the circular arc center, the horizontal chord $\tilde{h}$, and the traveltime $t_{\text{circ}}$ are,

$$\tilde{z} = \Delta z_v \quad , \quad \tilde{h} = 2\Delta z_v \cot \theta_a \quad , \quad k\, t_{\text{circ}} = \operatorname{arccosh}\left(1 + \frac{\tilde{h}^2}{2\tilde{z}^2}\right) \quad . \tag{D15}$$

The depth level of the arc center is labeled "center level" in Figure 13. The chord $\tilde{h}$ represents the subsurface offset on the interface between the layer and the half-space. The vertical distance $\Delta z_v$ and the subsurface half-offset $\tilde{h}/2$ are shown in Figure 13.

The total traveltime reads,

$$t = t_{\text{circ}} + 2 t_{\text{lin}} \quad , \tag{D16}$$

which leads to the normalized value,

$$k\, t[\theta_a(h)] = \operatorname{arccosh}\left(1 + \frac{2k^2 \Delta z_v^2 \cot^2 \theta_a}{v_a^2}\right) + \frac{2k\, z_h}{v_a \cos \theta_a} \quad , \tag{D17}$$

and simplifies to,

$$k\, t[\theta_a(h)] = \operatorname{arccosh}\left(1 + 2\cot^2 \theta_a\right) + \frac{2}{\cos \theta_a} \frac{z_h}{\Delta z_v} \quad , \tag{D18}$$

where $\theta_a(h)$ is given by equation D9. The lower sign (minus) in equation D9 corresponds to a caustic-free diving ray, and the upper sign (plus) – to a diving ray with the caustic. The constant horizontal slowness of the ray trajectory reads,

$$p_h = \frac{\sin \theta_a}{v_a} \quad . \tag{D19}$$

To find the geometric spreading, we need the first and second derivatives of the traveltime wrt the offset. For this, we apply the chain rule,

$$\frac{dt[\theta_a(h)]}{dh} = \frac{dt}{d\theta_a} \frac{d\theta_a}{dh} \quad , \quad \frac{d^2 t[\theta_a(h)]}{dh^2} = \frac{d^2 t}{d\theta_a^2} \left(\frac{d\theta_a}{dh}\right)^2 + \frac{dt}{d\theta_a} \frac{d^2 \theta_a}{dh^2} \quad . \tag{D20}$$

The ray trajectory equation reads,

$$\begin{array}{lll} \text{for} & -h/2 < x_1 < -\Delta z_v \cot \theta_a \ , & x_3 = (h/2 + x_1)\cot \theta_a \quad , \\ \text{for} & -\Delta z_v \cot \theta_a < x_1 < +\Delta z_v \cot \theta_a \ , & x_3 = z_h + \sqrt{\rho^2 - x_1^2} - \Delta z_v \quad , \\ \text{for} & +\Delta z_v \cot \theta_a < x_1 < +h/2 & , \quad x_3 = (h/2 - x_1)\cot \theta_a \quad . \end{array} \tag{D21}$$

where $x_1$ and $x_3$ are Cartesian coordinates of the path. Due to symmetry, the maximum penetration depth corresponds to vanishing $x_1$,

$$z_{\max} = z_h + \Delta z_v \frac{1 - \sin \theta_a}{\sin \theta_a} \quad . \tag{D22}$$

In constant gradient models, geometric spreading is equal to parameter $\sigma$, where for any 1D medium, $\sigma = h/p_h$. This is however, not the case for the given velocity model, i.e., $L_{GS} \neq \sigma$, because at the interface, the velocity gradient is discontinuous, and the second derivative of the velocity represents a delta-function, $k\delta(x_3 - z_h)$. Applying the formulae for geometric spreading, we found that for both caustic-free and caustic-generating trajectories, corresponding to the same offset, the following identity holds,

$$L_{GS} / \sigma = \sqrt[4]{1 - h_{min}^2 / h^2} \quad , \quad \sigma = h / p_h \qquad (D23)$$

In particular, for $h = h_{min}$, geometric spreading vanishes. For this particular case, the caustic happens at the endpoint of the ray path (at the receiver).

For the velocity model used in Figure 13, with $v_a = 1\,\text{km/s}$, $k = 1\,\text{s}^{-1}$, $z_h = 3\,\text{km}$, $\Delta z_v = 1\,\text{km}$, we consider an offset that exceeds the minimum value by 20%, $h / h_{min} = 1.2$. For two diving rays that co-exist with this offset (deep and shallow), and for the diving ray with a minimum offset $h_{min} = 4\sqrt{3}\,\text{km} = 6.9282032\,\text{km}$, we compute analytically a number of characteristics and summarize them in Table 10. All values are normalized (unitless). For a smoothed model with continuous velocity gradient, we compute the corresponding value numerically, with the use of the finite element method (Example 3), and list the results in Table 11.

The plots in Figure 14 are related to the unsmooth model (with a discontinuous velocity gradient). The two rays – with and without caustics, corresponding to the same offset $h = 1.2\,h_{min}$, are plotted in Figure 14b. In Figure 14c, we plot the caustic-free ray shown by a blue line in Figure 14b, and its two paraxial rays. All three rays start at the same source point and have slightly different offsets. The take-off angles of the paraxial ray are one degree less and one degree more than that of the central ray. As we see, the rays do not intersect each other.

Page 44 of 100

In Figure 14d, we plot the ray with a caustic shown by a red line in Figure 14b, and its two paraxial rays, with the take off-angles differing by $\pm 1°$ from that of the central ray. Again, all three rays start at the same source and arrive to different destinations. As we see, the rays do intersect each other. Figure 14e is a zoom of Figure 14d that shows the intersection of the two paraxial rays with the central ray and with each other. Figure 14f is yet one more zoom of Figure 14d; it shows that the three rays (central and two paraxial) intersect each other at different points.

Figure 14g shows the ray of the minimum offset that still allows a diving ray. Its take-off angle accepts the critical value $\theta_c$. For all smaller offsets, the only feasible trajectory is the straight line connecting the source and receiver on the earth's surface.

Since ray characteristic functions differ very slowly near the minimum offset point, in this case the take-off angles of the two paraxial rays, shown in Figure 14g, differ by $\pm 3°$ from that of the central rays (rather than $\pm 1°$ in the previous figures). Rays with take-off angles smaller than that of the critical ray (with the minimum offset), like the ray shown by a blue line in Figure 14b, are caustic-free. Rays with take-off angles exceeding that of the critical ray, like the ray shown by a red line in Figure 14b, have a caustic. Figure 14h is a zoom of Figure 14g. In Figure 14h, the "blue" paraxial ray (whose take-off angle exceeds that of the critical ray) intersects the central ray, while the "red" paraxial ray (whose take-off angle is below that of the central ray) does not. We also see that for the take-off angle difference of $3°$, the intersection is close to the destination point. For an infinitesimal difference, the caustic occurs at the destination (receiver). That is why the geometric spreading is zero for the critical ray (while parameter $\sigma = h / p_h$ is not).

**APPENDIX E. JACOBIAN IN THE SIMPLE**

# CAUSTIC-GENERATING VELOCITY MODEL

A vanishing Jacobian, determinant of the transform matrix between the ray coordinates (RC) and Cartesian coordinates, is the caustic criterion. For the simple unsmooth caustic-generating velocity model studied in Appendix D (a constant velocity layer over a constant velocity gradient half-space), we can ignore the 3D effects, and consider only two Cartesian coordinates, $x_1$ and $x_3$. There are also two ray coordinates (RC). Let $\theta_a$ be the take-off angle of the central ray. The first ray coordinate is $\gamma$, where $\omega = \theta_a + \gamma$ is the take-off angle of the paraxial ray. The second ray coordinate is the traveltime along the central ray. The ray Jacobian reads,

$$J_{\text{ray}} = \det \mathbf{Q} \, , \quad \mathbf{Q} = \begin{bmatrix} \dfrac{\partial x_1}{\partial \gamma} & \dfrac{\partial x_1}{\partial t} \\ \dfrac{\partial x_3}{\partial \gamma} & \dfrac{\partial x_3}{\partial t} \end{bmatrix} \, . \tag{E1}$$

The ray trajectory can be split into four parts:

- A straight line in the constant velocity layer, between the source on the earth's surface and the medium interface, $x_3 = z_h$.

- The first half of the circular arc in the constant gradient half-space, between the medium interface and the point of maximum penetration depth, $x_3 = z_{\max}$.

- The second half of the circular arc in the constant gradient half-space, between the point of maximum depth and the medium interface

- A straight line in the constant velocity layer, between the medium interface and the receiver on the surface

The trajectory is symmetric wrt the vertical line of the source-receiver midpoint.

Before studying the Jacobian, one can compute the traveltimes of the paraxial ray.

For the straight-line path,

$$t_{\text{lin}} = \frac{z_h}{v_a \cos \omega} \quad . \tag{E2}$$

For the half-arc (Červený, 2000),

$$k\, t_{\text{arc}} = \text{arccosh}\left[1 + \frac{\Delta x^2 + (\tilde{z}_1 - \tilde{z}_2)^2}{2\tilde{z}_1 \tilde{z}_2}\right] \quad, \tag{E3}$$

where $\Delta x$ is the horizontal distance for the half-arc (i.e., the half-chord), and $\tilde{z}_1$, $\tilde{z}_2$ are depths of the half-arc endpoints, measured from the arc center level,

$$\Delta x = \rho_\omega \cos \omega \,, \quad \tilde{z}_1 = \rho_\omega \sin \omega \,, \quad \tilde{z}_2 = \rho_\omega \,, \quad \rho_\omega = \Delta z_h / \sin \omega \quad, \tag{E4}$$

where $\rho_\omega$ is the radius of the paraxial arc. The half-arc traveltime becomes,

$$k\, t_{\text{arc}} = \text{arccosh}\left[1 + \frac{\cos^2 \omega + (1 - \sin \omega)^2}{2 \sin \omega}\right] = \text{arccosh}\frac{1}{\sin \omega} \quad . \tag{E5}$$

An alternative formula for half-arc traveltime follows from equation D18,

$$k\, t_{\text{arc}} = \frac{1}{2}\text{arccosh}\left(1 + 2\cot^2 \omega\right) \quad . \tag{E6}$$

The right sides of equations E5 and E6 are identical. To make sure, one can use two auxiliary formulae,

$$\cosh \frac{x}{2} = \sqrt{\frac{\cosh x + 1}{2}} \ , \qquad \sqrt{1 + \cot^2 \omega} = \frac{1}{\sin \omega} \ , \qquad 0 < \omega \leq \pi/2 \qquad . \qquad \text{(E7)}$$

For the central ray, $\gamma = 0$, and $\omega = \theta_a$. The offsets of the central and paraxial rays are $h$ and $h_\gamma$, respectively,

$$\begin{aligned} h &= 2z \tan \theta_a + 2\Delta z_v \cot \theta_a \ , \\ h_\gamma &= 2z \tan \omega + 2\Delta z_v \cot \omega \ . \end{aligned} \qquad \text{(E8)}$$

Both rays start at the same surface point with the lateral coordinate $x_1 = -h/2$. We split the whole ray path into four time intervals as mentioned above and shown in Figure 17,

$$\begin{aligned} &\bullet \qquad 0 \leq t \leq t_{\text{lin}} \ , \\ &\bullet \qquad t_{\text{lin}} \leq t \leq t_{\text{lin}} + t_{\text{arc}} \ , \\ &\bullet \qquad t_{\text{lin}} + t_{\text{arc}} \leq t \leq t_{\text{lin}} + 2t_{\text{arc}} \ , \\ &\bullet \qquad t_{\text{lin}} + 2t_{\text{arc}} \leq t \leq 2t_{\text{lin}} + 2t_{\text{arc}} \ . \end{aligned} \qquad \text{(E9)}$$

For each time interval, we need to find $x_1$ and $x_3$ as functions of $\gamma$ and $t$.

On the first interval,

$$x_1 = -h/2 + v_a t \sin \omega \ , \quad x_3 = v_a t \cos \omega \qquad . \qquad \text{(E10)}$$

At the end of the first interval, the Cartesian coordinates read,

$$x_1^{(1)} = -h/2 + z_h \tan \omega \ , \quad x_3^{(1)} = z_h \qquad . \qquad \text{(E11)}$$

Next we study the second time interval (the first half of the arc). At the end of the second interval, the coordinates read,

$$x_1^{(2)} = -h/2 + z_h \tan \omega + \Delta z_v \cot \omega , \quad x_3^{(2)} = z_h - \Delta z_v + \rho \quad . \tag{E12}$$

It is convenient to introduce the time remaining to the end of this interval,

$$t_{\text{rem}} = t_{\text{lin}} + t_{\text{arc}} - t , \quad dt_{\text{rem}} / dt = -1 \quad , \tag{E13}$$

where $t$ is the current time. Note that the remaining time $t_{\text{rem}}$ is a positive value. We relate the remaining time with the central angle $\alpha$ to the left of the vertical line (see Figure 17). This angle is assumed negative for the second interval. Coordinates of a point on the arc are,

$$x_1 = x_1^{(2)} + \rho_\omega \sin \alpha , \quad x_3 = x_3^{(2)} - \rho_\omega (1 - \cos \alpha) = x_3^{(2)} - 2\rho_\omega \sin^2 \frac{\alpha}{2} \quad , \tag{E14}$$

where,

$$-(\pi/2 - \omega) \leq \alpha \leq 0 \quad . \tag{E15}$$

Angle $\alpha$ depends on the ray coordinate $\gamma$ (or $\omega$ which is $\theta_a + \gamma$) and on the remaining time $t_{\text{rem}}$ defined in equation E13,

$$k t_{\text{rem}} = \operatorname{arccosh} \left[ 1 + \frac{\Delta x^2 + (\tilde{z}_1 - \tilde{z}_2)^2}{2 \tilde{z}_1 \tilde{z}_2} \right] \quad . \tag{E16}$$

In this case,

$$\Delta x = \rho_\omega \sin \alpha , \quad \tilde{z}_1 = \rho_\omega \cos \alpha , \quad \tilde{z}_2 = \rho_\omega \quad , \tag{E17}$$

which leads to,

$$k\, t_{rem} = \text{arccosh}\left[1 + \frac{\sin^2 \alpha + (1-\cos \alpha)^2}{2\cos \alpha}\right] = \text{arccosh}\frac{1}{\cos \alpha} \quad . \quad (E18)$$

Equation E18 relates $t_{rem}(\alpha)$, and we need to invert it for $\alpha(t_{rem})$. Cosine is an even function. Since we agreed to consider angle $\alpha$ negative within the second ray path interval, this results in,

$$\alpha = -\text{arccos}\frac{1}{\cosh k\, t_{rem}} \quad . \quad (E19)$$

where $t_{rem}$ is given in equation E13.

Our third interval is the second half of the circular arc. We treat it in the same way, but the central angle $\alpha$ is now to the right of the vertical line, and this angle is assumed positive. Equation L14 for the second interval holds for the third interval as well, but the range is now,

$$0 \leq \alpha \leq \pi/2 - \omega \quad . \quad (E20)$$

The remaining time is now counted from the beginning of the third interval,

$$t - (t_{lin} + t_{arc}) = t_{rem}, \quad dt_{rem}/dt = +1 \quad . \quad (E21)$$

The function on the right side of equation E19 is even wrt $t_{rem}$, so the sign of $t_{rem}$ does not matter; however, we agree to consider $\alpha$ positive within the third interval. This leads to,

$$\alpha = +\text{arccos}\frac{1}{\cosh k\, t_{rem}} \quad . \quad (E22)$$

We will need the derivative of the function wrt the absolute time (measured from the source point)

$$\frac{d\alpha}{dt} = \frac{k}{\cosh kt_{\text{rem}}} \quad . \quad (E23)$$

Recall that $dt_{\text{rem}}/dt = -1$ for the second interval, and $dt_{\text{rem}}/dt = +1$. Equation E23 is valid for both the second and third intervals, where the derivative $d\alpha/dt$ is positive. At the end of the third interval, the coordinates are,

$$x_1^{(3)} = -h/2 + z_h \tan \omega + 2\Delta z_v \cot \omega , \quad x_3^{(3)} = z_h \quad . \quad (E24)$$

Eventually, on the fourth interval, the Cartesian coordinates of a ray point are,

$$x_1 = x_1^{(3)} + v_a (t - t_{\text{lin}} - 2t_{\text{arc}}) \sin \omega , \quad x_3 = z_h - v_a (t - t_{\text{lin}} - 2t_{\text{arc}}) \cos \omega \quad . \quad (E25)$$

At the end of the fourth interval (at the receiver of the paraxial ray), the coordinates are,

$$x_1^{(4)} = -h/2 + 2z_h \tan \omega + 2\Delta z_v \cot \omega = -h/2 + h_\omega , \quad x_3^{(4)} = 0 \quad . \quad (E26)$$

The second column of the transform matrix represents the velocity components. On the first interval,

$$v_1 = v_a \sin \omega , \quad v_3 = v_a \cos \omega , \quad v_1 > 0 , \quad v_3 > 0 \quad . \quad (E27)$$

On the second and the third intervals,

$$v_1 = +\rho_\omega \cos \alpha \frac{\partial \alpha}{\partial t} , \quad v_3 = -\rho_\omega \sin \alpha \frac{\partial \alpha}{\partial t} , \quad (E28)$$

where,

$$\text{Interval 2:} \quad \alpha < 0, \quad \frac{\partial \alpha}{\partial t} > 0, \quad v_1 > 0, \quad v_3 > 0,$$
$$\text{Interval 3:} \quad \alpha > 0, \quad \frac{\partial \alpha}{\partial t} > 0, \quad v_1 > 0, \quad v_3 < 0.$$
(E29)

On the fourth interval,

$$v_1 = v_a \sin \omega, \quad v_3 = -v_a \cos \omega, \quad v_1 > 0, \quad v_3 < 0 \quad . \quad (E30)$$

The Jacobian should be established for $\gamma = 0$, i.e., transformation derivatives should be computed along the central ray. With the above workflow, we obtain analytical expressions for Jacobian $J_{\text{ray}}(t)$ for all four intervals of the ray path. These formulae are too lengthy to be explicitly presented in this study, so we plot the graphs. The range is $0 \leq t \leq 2t_{\text{lin}} + 2t_{\text{arc}}$ of the central ray. In Figure 18, the Jacobian vs. traveltime is plotted for the three rays studied above: a) the caustic-free ray with the offset $h = 1.2 h_{\min}$, b) the ray with the caustic and the same offset (multi-arrival), and c) the ray with the minimum offset $h_{\min}$ that still allows the diving ray. In the latter case, the caustic occurs at the endpoint (at the receiver location). For the model considered, in all three cases the Jacobian proves to be discontinuous at the interface between the constant velocity layer and the constant velocity gradient half-space. For the smooth model considered in the numerical Example 3, the Jacobian is continuous. The caustic detected is the first-order caustic (line). In this simple analysis, we ignored dimension $x_2$ normal to the plane of the ray trajectory, assuming that there are no intersections of the central and paraxial rays with a vanishing (out of the central ray path plane) component $x_2$ of the normal shift (and thus, considering a 2D problem instead of 3D).

# APPENDIX F.

## SADDLE POINT TRAVELTIME IN THE SIMPLE CAUSTIC-GENERATING MEDIUM

As mentioned, the saddle point in the simple caustic-generating medium can be viewed as the maximum traveltime for one DoF, and the minimum for all other DoF. In this discussion, we consider a diving ray of a given surface offset $h$ in the simple caustic-generating medium (Example 3 in the body of the paper) as a path that can be described by two DoF only: a) the subsurface horizontal offset $\tilde{h}$ at the interface between the constant velocity layer and the half-space layer with the constant velocity gradient, and b) the maximum penetration depth $z$, see Figure 15. The stationary diving ray traveltime for this scheme is always a minimum for $\tilde{h}$, but may be both a minimum and a maximum for $z$.

To demonstrate this, consider a two-way reflection trajectory, rather than a diving ray. (For the diving ray solution, this reflection path is non-stationary.) Let $z$ be the depth of the flat reflector, $z > z_h$, where $z_h$ is the thickness of the constant velocity layer. The ray path is symmetric, so only one half of it is shown in Figure 15. The two-way traveltime $t$ reads,

$$\frac{kt}{2} = \frac{\sqrt{(h-\tilde{h})^2 + 4z_h^2}}{2\Delta z_v} + \operatorname{arccosh}\left[1 + \frac{\tilde{h}^2 + 4(z-z_h)^2}{8\Delta z_v (z - z_h + \Delta z_v)}\right] \quad , \tag{F1}$$

where the unknown parameter $\tilde{h}$ is the subsurface half-offset at the interface between the constant velocity and the constant gradient layers, $\tilde{h} < h$. This parameter can be defined from the stationarity of the reflection traveltime,

$$\frac{k}{2}\frac{dt}{dh} = \frac{\tilde{h}}{\sqrt{\left[\tilde{h}^2 + 4(z-z_h)^2\right]\left[\tilde{h}^2 + 4(2\Delta z_v + z - z_h)^2\right]}} - \frac{h-\tilde{h}}{2\Delta z_v \sqrt{(h-\tilde{h})^2 + 4z_h^2}} = 0 \quad , \quad \text{(F2)}$$

where $v_a = k\Delta z_v$. This equation has a single real root,

$$\tilde{h}(z) = \frac{h+P}{3} + \frac{h^2}{3P} - 4\frac{2\Delta v_z\, z + (z-z_h)^2}{P} \quad , \quad \text{(F3)}$$

where,

$$\begin{aligned}
P &= \sqrt[3]{h^3 + Q + R} \quad , \\
R &= 36h\left[\Delta z_v(2z - 3h) + (z-z_h)^2\right] \quad , \\
Q &= \sqrt{\left[24z\Delta z_v + 12(z-z_h)^2 - h^2\right]^3 + \left(h^3 + R\right)^2} \quad .
\end{aligned} \quad \text{(F4)}$$

This value of the subsurface offset $a$ yields a minimum traveltime $t(z)$ for any depth $z$ of the reflector. A minimum can be confirmed by analysis of the second derivative, $d^2t/d\tilde{h}^2$ for the stationary point where the first derivative $dt/d\tilde{h}$ vanishes. After elimination of the subsurface offset $\tilde{h}$ (it is now a function of the reflector depth $z$ rather than an independent parameter), the ray path has a single DoF $z$, and we can identify the stationary points with the vanishing derivative, $dt\left[\tilde{h}(z), z\right]/dz = 0$. We just plot the graph and see the zeros.

We analyze the incidence/reflection angle $\theta_a(z)$, to make sure that the stationary path corresponds to $\theta_a = 90°$ (the diving ray). The conservation of the horizontal slowness along the ray reads,

$$p_h = \frac{\sin\theta_a}{v_a + k(z - z_h)} = \frac{\sin\theta_b}{v_a} \quad , \quad \sin\theta_b = \frac{h - \tilde{h}}{\sqrt{(h - \tilde{h})^2 + 4z_h^2}} \quad , \tag{F5}$$

where $\theta_b$ is the departure/arrival angle at the surface. This leads to,

$$\sin\theta_a(z) = \frac{z + \Delta z_v - z_h}{\Delta z_v} \cdot \frac{h - \tilde{h}}{\sqrt{(h - \tilde{h})^2 + 4z_h^2}} \quad . \tag{F6}$$

If the reflector is above the interface between the layers, the two-way traveltime reads,

$$\frac{kt}{2} = \frac{\sqrt{h^2 + 4z^2}}{2\Delta z_v} \quad , \quad \sin\theta_a(z) = \sin\theta_b(z) = \frac{h}{\sqrt{h^2 + 4z^2}} \quad . \tag{F7}$$

Next, we assume that the reflector depth is a varying parameter, and we plot the traveltime (Figure 16a), its derivative $dt/dz$ (Figure 16b), and the reflection angle $\theta_a$ (Figure 16c) vs. the reflector depth $z$. The dashed lines correspond to the reflector above the interface between the constant velocity layer and the constant gradient layer, while the solid lines correspond to the reflector below this interface.

Graphs for the traveltime $t(z)$ and $dt(z)/dz$ show that for the given unsmooth model, there are two minima: at $z_o = 0$, $t_{\min} = 8.3138439$ s and at $z_2 = 5.3787482$, $t_{\min} = 10.057414$ s, and a maximum between them: at $z_1 = 3.3653059$, $t_{\max} = 10.474043$ s. A minimum at zero depth is trivial and corresponds to the straight line connecting the source and receiver.

A graph for the incidence/reflection angles shows that for all three "stationary depths", $z_o, z_1, z_2$, these angles are $90^o$. This means that for both $z_1$ (maximum traveltime) and $z_2$ (minimum traveltime) there is no reflection - only a diving ray exists.

In this theoretical analysis, the path of the diving ray has only two DoF: the subsurface offset $\tilde{h}$ (on the interface between the two layers) and the maximum penetration depth $z$. Comparing this analytical scheme with the finite-element discretization, we assume that the maximum depth $z$ corresponds to the vertical coordinate of the central node of the path, while the subsurface offset $\tilde{h}$ corresponds to all other discrete DoF of the numerical implementation. The stationary traveltime in this example is always minimum vs. coordinate $\tilde{h}$, but may be a minimum or a maximum vs. coordinate $z$. A minimum vs. both $\tilde{h}$ and $z$ yields a true traveltime minimum, while a minimum vs. $\tilde{h}$ and a maximum vs. $z$ yields a saddle point solution.

## APPENDIX G. DIVING WAVES
## IN ELLIPSOIDAL ANISOTROPIC MODEL WITH TILTED GRADIENT

Our goal is to compute the trajectory of the diving ray in the symmetry plane of an ellipsoidal anisotropy. An ellipsoidal model can be considered a particular case of an orthorhombic model. An acoustic wave is considered, and all three intrinsic anellipticities vanish, $f = 1, \eta_1 = \eta_2 = \eta_3 = 0$, where,

Page 56 of 100

$$\eta_1 = \frac{\varepsilon_1 - \delta_1}{1 + 2\delta_1} , \qquad \eta_2 = \frac{\varepsilon_2 - \delta_2}{1 + 2\delta_2} ,$$

$$\eta_3 = \frac{\varepsilon_1 - \varepsilon_2 - \delta_3 (1 + 2\varepsilon_2)}{(1 + 2\varepsilon_2)(1 + 2\delta_2)} , \qquad f = 1 - v_{S1}^2 / v_P^2 ,$$

(G1)

which leads to,

$$\varepsilon_1 = \delta_1 , \quad \varepsilon_2 = \delta_2 , \quad \delta_3 = \frac{\varepsilon_1 - \varepsilon_2}{1 + 2\varepsilon_2} , \tag{G2}$$

or alternatively (e.g., Pouya and Chalhoub, 2007),

$$C_{12} = \sqrt{C_{11} C_{22}} , \quad C_{13} = \sqrt{C_{11} C_{33}} , \quad C_{23} = \sqrt{C_{22} C_{33}} , \quad C_{44} = C_{55} = C_{66} = 0 . \tag{G3}$$

The slowness surface represents an ellipsoid,

$$C_{11} p_1^2 + C_{22} p_2^2 + C_{33} p_3^2 = 1 , \tag{G4}$$

with the axial velocities,

$$A_v = \sqrt{C_{11}} = v_P \sqrt{1 + 2\varepsilon_2} , \quad B_v = \sqrt{C_{22}} = v_P \sqrt{1 + 2\varepsilon_1} , \quad C_v = \sqrt{C_{33}} = v_P . \tag{G5}$$

Further in this appendix, we consider the acoustic wave propagation in $x_1 x_3$ symmetry plane that corresponds to our numerical Example 5. According to equation G10 of Part II, the ray velocity reads,

$$v_{\text{ray}}(\mathbf{x}, \mathbf{r}) = \frac{A_v(\mathbf{x}) B_v(\mathbf{x}) C_v(\mathbf{x})}{\sqrt{B_v^2(\mathbf{x}) C_v^2(\mathbf{x}) r_1^2 + A_v^2(\mathbf{x}) C_v^2(\mathbf{x}) r_2^2 + A_v^2(\mathbf{x}) B_v^2(\mathbf{x}) r_3^2}} . \tag{G6}$$

In the symmetry plane,

$$r_1 = \sin \theta_{\text{ray}} , \quad r_2 = 0 , \quad r_3 = \cos \theta_{\text{ray}} , \tag{G7}$$

and the ray velocity simplifies to

$$v_{\text{ray}}(\mathbf{x}, \mathbf{r}) = \frac{A_v(\mathbf{x}) C_v(\mathbf{x})}{\sqrt{C_v^2(\mathbf{x}) \sin^2 \theta_{\text{ray}} + A_v^2(\mathbf{x}) \cos^2 \theta_{\text{ray}}}} . \tag{G8}$$

We consider a particular case, where the vertical velocity $C_v(\mathbf{x})$ is equal to the isotropic reference velocity $v(\mathbf{x})$, and the horizontal velocity is proportional to the vertical velocity with the constant (coordinate-independent) coefficient,

$$A_v(\mathbf{x}) = \sqrt{1+2\varepsilon_2}\, C_v(\mathbf{x}) \equiv \lambda C_v(\mathbf{x}) \qquad . \qquad (G9)$$

We assume here that $\varepsilon_2$ is a positive value, thus $\lambda > 1$. With this equation, the ray velocity simplifies to,

$$v_{\text{ray}}(\mathbf{x}, \theta_{\text{ray}}) = \frac{\lambda v(\mathbf{x})}{\sqrt{\sin^2 \theta_{\text{ray}} + \lambda^2 \cos^2 \theta_{\text{ray}}}} \qquad . \qquad (G10)$$

The traveltime becomes,

$$t = \int_S^R \frac{ds}{v_{\text{ray}}} = \int_S^R \frac{\sqrt{1+z'^2}\, dx}{v_{\text{ray}}} \qquad , \qquad (G11)$$

where $z(x)$ is the ray path. Note that the derivative $z'(x)$ is related to the ray angle,

$$z' = \cot \theta_{\text{ray}} \;\rightarrow\; \sin^2 \theta_{\text{ray}} = \frac{1}{1+z'^2} \;,\; \cos^2 \theta_{\text{ray}} = \frac{z'^2}{1+z'^2} \;, \qquad (G12)$$

where $\theta_{\text{ray}}$ is measured from the vertical line. Combining equations G11 and G12, we obtain,

$$v_{\text{ray}}(x, z, z') = \lambda v(x, z) \frac{\sqrt{1+z'^2}}{\sqrt{1+\lambda^2 z'^2}} \qquad . \qquad (G13)$$

Introduction of this result into equation G11 leads to,

$$t = \int_S^R \frac{ds}{v_{\text{ray}}} = \int_S^R \frac{\sqrt{1+\lambda^2 z'^2}}{\lambda v(x,z)} dx \qquad (G14)$$

The function is not parametric here, but just a function of a single variable. Applying the Euler-Lagrange equation,

$$\frac{d}{dx} \frac{\partial}{\partial z'} \frac{\sqrt{1+\lambda^2 z'^2}}{\lambda v(x,z)} = \frac{\partial}{\partial z} \frac{\sqrt{1+\lambda^2 z'^2}}{\lambda v(x,z)} \qquad (G15)$$

we obtain the second-order ODE,

$$\frac{d}{dx} \frac{\lambda z'}{v(x,z)\sqrt{1+\lambda^2 z'^2}} = -\frac{\sqrt{1+\lambda^2 z'^2}}{\lambda v^2(x,z)} \frac{\partial v(x,z)}{\partial z} \qquad (G16)$$

It is convenient to introduce the scaled vertical distance, $\hat{z} = \lambda z$. With the chain rule, the vertical component of the reference velocity gradient reads,

$$\frac{\partial v(x, \hat{z}/\lambda)}{\partial \hat{z}} = \frac{\partial v(x,z)}{\partial z} \frac{dz}{d\hat{z}} = \frac{1}{\lambda} \frac{\partial v(x,\hat{z})}{\partial z} \qquad (G17)$$

and the ray path ODE G15 can be arranged as,

$$\frac{d}{dx} \frac{\hat{z}'}{v(x,\hat{z})\sqrt{1+\hat{z}'^2}} = -\frac{\sqrt{1+\hat{z}'^2}}{v^2(x,z)} \frac{\partial v(x,\hat{z})}{\partial \hat{z}} \qquad (G18)$$

The anisotropic factor $\lambda = \sqrt{1+2\varepsilon_2}$ is now hidden. The ray tracing algorithm can be arranged as follows:

- Rescale the velocity field $v(x,z)$ into $v(x,\hat{z})$, where $\hat{z} = \lambda z$

- Find the ray path $\hat{z} = \hat{z}(x)$ for the isotropic velocity field $v(x, \hat{z})$

- Rescale the vertical coordinate back to $z = \hat{z}/\lambda$

In particular, this means that for a reference velocity with the constant gradient (vertical or tilted), the circular trajectory in the reference velocity field becomes elliptic for the considered type of anisotropy. Applying equation B4, we obtain the parameters of the ellipse: the horizontal and vertical radii $A_e$ and $B_e$, the eccentricity $m_e$ and the maximum penetration depth $z_{max}$,

$$A_e = \frac{\sqrt{4v_a^2 \lambda^2 + k_3^2 h^2}}{2k_3} \quad , \quad B_e = \frac{\sqrt{4v_a^2 \lambda^2 + k_3^2 h^2}}{2k_3 \lambda} \quad ,$$
$$m_e = \frac{\sqrt{\lambda^2 - 1}}{\lambda} \quad , \quad z_{max} = B_e - \frac{v_a}{k_3} \quad .$$
(G19)

The take-off / arrival angles, $\theta_a$, $\theta_b$, and the arclength $s$ of the ray path are,

$$\theta_a = \arctan \frac{2\lambda^2 v_a}{k_3 h} \quad , \quad \theta_b = \pi - \theta_a \quad , \quad s = \frac{B_e^2}{A_e} \int_{\theta_a}^{\theta_b} \left(1 - m_e^2 \sin^2 \theta\right)^{-3/2} d\theta \quad .$$
(G20)

Coordinates $x$ and $z$ and the ray velocity can be expressed through the angle,

$$x(\theta) = \frac{A_e \cos \theta}{\sqrt{1 - m_e^2 \sin^2 \theta}} \quad ,$$
$$z(\theta) = \frac{\sqrt{1 - m_e^2} B_e \sin \theta}{\sqrt{1 - m_e^2 \sin^2 \theta}} - \frac{v_a}{k_3} \quad , \quad v_{ray}(\theta) = \frac{v_a + k_1 x(\theta) + k_3 z(\theta)}{\sqrt{1 - m_e^2 \sin^2 \theta}} \quad .$$
(G21)

This makes it possible to find the traveltime and sigma analytically,

$$d\tau = \frac{ds}{v_{ray}} \quad , \quad d\sigma = v_{ray} \, ds \quad , \quad ds = \frac{B_e^2}{A_e} \left(1 - m_e^2 \sin^2 \theta\right)^{-3/2} d\theta \quad ,$$
(G22)

by integrating equation G22.

# LIST OF TABLES

Table 1. Nodal locations and orientations for the constant vertical velocity gradient path.

Table 2. Endpoint location Hessian, s/km$^2$ for the constant vertical velocity gradient model. The mixed *RS* block is highlighted in yellow.

Table 3. Accuracy of the constant vertical velocity gradient path.

Table 4. Nodal locations and orientations for the constant tilted velocity gradient path.

Table 5. Endpoint location Hessian, s/km$^2$ for the constant tilted velocity gradient model. The mixed *RS* block is highlighted in yellow.

Table 6. Accuracy of the constant tilted velocity gradient path.

Table 7. Nodal locations and orientations for the conic velocity model path.

Table 8. Endpoint location Hessian, s/km$^2$ for the conic velocity model. The mixed *RS* block is highlighted in yellow.

Table 9. Accuracy of the conic velocity model path.

Table 10. Normalized characteristics of deep, shallow, and minimum offset waves in the simple, unsmooth, caustic-generating medium (computed analytically).

Table 11. Normalized characteristics of deep, shallow, and minimum offset waves in the simple, smooth, caustic-generating medium (computed numerically).

Table 12. Kinematic characteristics of ray paths for the gas-cloud velocity model.

Table 13. Dynamic characteristics of ray paths for the gas-cloud velocity model.

Table 14. Nodal locations and orientations of the path, for ellipsoidal model with constant velocity gradient.

Table 15. Endpoint location Hessian, s/km$^2$ for ellipsoidal velocity model with constant gradient. The mixed *RS* block is highlighted in yellow.

Table 16. Accuracy of the ray path for ellipsoidal velocity model with constant gradient.

Table 17. Conversion velocity and geometric spreading (GS) for ellipsoidal model with constant velocity gradient.

Table 18. Nodal locations and orientations of the path, for ellipsoidal model with varying velocity gradient.

Table 19. Endpoint location Hessian, s/km$^2$ for ellipsoidal velocity model with varying velocity gradient. The mixed $RS$ block is highlighted in yellow.

Table 20. Kinematic and dynamic characteristics for ellipsoidal model with constant velocity gradient.

Table 1. Nodal locations and orientations for the constant vertical velocity gradient path.

| node | location, km | | orientation | |
|---|---|---|---|---|
| | $x_1$ | $x_3$ | $r_1$ | $r_3$ |
| 0 | –5 | 0 | 0.371389 | +0.928477 |
| 1 | –4.64828 | 0.719093 | 0.504923 | +0.863164 |
| 2 | –4.19385 | 1.37810 | 0.627299 | +0.778779 |
| 3 | –3.64676 | 1.96247 | 0.735812 | +0.677186 |
| 4 | –3.01908 | 2.45928 | 0.828067 | +0.560629 |
| 5 | –2.32469 | 2.85755 | 0.902025 | +0.431684 |
| 6 | –1.57893 | 3.14849 | 0.956051 | +0.293200 |
| 7 | –0.798285 | 3.32567 | 0.988952 | +0.148238 |
| 8 | 0 | 3.38516 | 1 | 0 |
| 9 | +0.798285 | 3.32567 | 0.988952 | –0.148238 |
| 10 | +1.57893 | 3.14849 | 0.956051 | –0.293200 |
| 11 | +2.32469 | 2.85755 | 0.902025 | –0.431684 |
| 12 | +3.01908 | 2.45928 | 0.828067 | –0.560629 |
| 13 | +3.64676 | 1.96247 | 0.735812 | –0.677186 |

| 14 | +4.19385 | 1.37810 | 0.627299 | –0.778779 |
| 15 | +4.64828 | 0.719093 | 0.504923 | –0.863164 |
| 16 | +5 | 0 | 0.371389 | –0.928477 |

Table 2. Endpoint location Hessian, s/km² for the constant vertical velocity gradient model. The mixed $RS$ block is highlighted in yellow.

| DoF | $x_{S,1}$ | $x_{S,2}$ | $x_{S,3}$ | $x_{R,1}$ | $x_{R,2}$ | $x_{R,3}$ |
|---|---|---|---|---|---|---|
| $x_{S,1}$ | $-1.58355 \times 10^{-2}$ | 0 | $+6.72353 \times 10^{-3}$ | $+1.60001 \times 10^{-2}$ | 0 | $+6.41185 \times 10^{-3}$ |
| $x_{S,2}$ | 0 | $+1.85695 \times 10^{-2}$ | 0 | 0 | $-1.85695 \times 10^{-2}$ | 0 |
| $x_{S,3}$ | $+6.72353 \times 10^{-3}$ | 0 | $+2.67219 \times 10^{-1}$ | $-6.41185 \times 10^{-3}$ | 0 | $-2.56947 \times 10^{-3}$ |
| $x_{R,1}$ | $+1.60001 \times 10^{-2}$ | 0 | $-6.41185 \times 10^{-3}$ | $-1.58355 \times 10^{-2}$ | 0 | $-6.72353 \times 10^{-3}$ |
| $x_{R,2}$ | 0 | $-1.85695 \times 10^{-2}$ | 0 | 0 | $+1.85695 \times 10^{-2}$ | 0 |
| $x_{R,3}$ | $+6.41185 \times 10^{-3}$ | 0 | $-2.56947 \times 10^{-3}$ | $-6.72353 \times 10^{-3}$ | 0 | $+2.67219 \times 10^{-1}$ |

Table 3. Accuracy of the constant vertical velocity gradient path.

| Characteristic | Notation | Exact | Numerical | Rel. error |
|---|---|---|---|---|
| Take-off angle | $\theta_a$, rad | 0.38050638 | 0.38050447 | $-5.00 \cdot 10^{-6}$ |
| Max. depth | $z_{max}$, km | 3.3851648 | 3.3851644 | $-1.22 \cdot 10^{-7}$ |
| Path arclength | $s$, km | 12.8198151 | 12.8198146 | $-4.18 \cdot 10^{-8}$ |
| Traveltime | $t$, s | 3.29446229274 | 3.29446229265 | $+2.69 \cdot 10^{-11}$ |
| Horiz. slowness | $p_h$, s/km | 0.18569534 | 0.18569536 | $+9.28 \cdot 10^{-8}$ |
| Sigma | $\sigma$, km²/s | 53.851648 | 53.851643 | $-1.02 \cdot 10^{-7}$ |
| Geom. spreading | $L_{GS}$, km²/s | 53.851648 | 53.851643 | $-1.02 \cdot 10^{-7}$ |

Table 4. Nodal locations and orientations for the constant tilted velocity gradient path.

| node | location, km | | orientation | |
|---|---|---|---|---|
| | $x_1$ | $x_3$ | $r_1$ | $r_3$ |
| 0 | –5 | 0 | 0.6 | +0.8 |
| 1 | –4.53275 | 0.553099 | 0.688496 | +0.725240 |
| 2 | –4.00467 | 1.04845 | 0.767752 | +0.640747 |
| 3 | –3.42285 | 1.47940 | 0.836704 | +0.547655 |
| 4 | –2.79508 | 1.84017 | 0.894427 | +0.447214 |
| 5 | –2.12981 | 2.12592 | 0.940147 | +0.340770 |
| 6 | –1.43596 | 2.13281 | 0.973249 | +0.229753 |
| 7 | –0.722828 | 2.45806 | 0.993290 | +0.115653 |
| 8 | 0 | 2.5 | 1 | 0 |
| 9 | +0.722828 | 2.45806 | 0.993290 | –0.115653 |
| 10 | +1.43596 | 2.13281 | 0.973249 | –0.229753 |
| 11 | +2.12981 | 2.12592 | 0.940147 | –0.340770 |
| 12 | +2.79508 | 1.84017 | 0.894427 | –0.447214 |
| 13 | +3.42285 | 1.47940 | 0.836704 | –0.547655 |
| 14 | +4.00467 | 1.04845 | 0.767752 | –0.640747 |
| 15 | +4.53275 | 0.553099 | 0. 688496 | –0.725240 |
| 16 | +5 | 0 | 0.6 | –0.8 |

Table 5. Endpoint location Hessian, s/km² for the constant tilted velocity gradient model. The mixed RS block is highlighted in yellow.

| DoF | $x_{S,1}$ | $x_{S,2}$ | $x_{S,3}$ | $x_{R,1}$ | $x_{R,2}$ | $x_{R,3}$ |
|---|---|---|---|---|---|---|
| $x_{S,1}$ | $+4.64522 \times 10^{-3}$ | 0 | $+5.94627 \times 10^{-2}$ | $+1.27945 \times 10^{-2}$ | 0 | $+9.60112 \times 10^{-3}$ |
| $x_{S,2}$ | 0 | $+2 \times 10^{-2}$ | 0 | 0 | $-2 \times 10^{-2}$ | 0 |
| $x_{S,3}$ | $+5.94627 \times 10^{-2}$ | 0 | $+2.05865 \times 10^{-1}$ | $-9.60208 \times 10^{-3}$ | 0 | $-7.20550 \times 10^{-3}$ |
| $x_{R,1}$ | $+1.27945 \times 10^{-2}$ | 0 | $-9.60208 \times 10^{-3}$ | $-1.37592 \times 10^{-2}$ | 0 | $+5.04855 \times 10^{-3}$ |
| $x_{R,2}$ | 0 | $-2 \times 10^{-2}$ | 0 | 0 | $+2 \times 10^{-2}$ | 0 |
| $x_{R,3}$ | $+9.60112 \times 10^{-3}$ | 0 | $-7.20550 \times 10^{-3}$ | $+5.04855 \times 10^{-3}$ | 0 | $+6.65535 \times 10^{-2}$ |

Table 6. Accuracy of the constant tilted velocity gradient path.

| Characteristic | Notation | Exact | Numerical | Rel. error |
|---|---|---|---|---|
| Departure angle | $\theta_S$, rad | 0.64350111 | 0.64350083 | $-4.35 \cdot 10^{-7}$ |
| Arrival angle | $\theta_R$, rad | 2.49809154 | 2.49809195 | $+1.62 \cdot 10^{-7}$ |
| Max depth | $z_{max}$, km | 2.5 | 2.499999937 | $-2.54 \cdot 10^{-8}$ |
| Path arclength | $s$, km | 11.59119023 | 11.59119015 | $-6.62 \cdot 10^{-9}$ |
| Traveltime | $t$, s | 2.840309249232 | 2.840309924242 | $-3.26 \cdot 10^{-12}$ |
| Horiz. slowness at S | $p_{h,S}$, s/km | 0.3 | 0.29999989 | $-3.73 \cdot 10^{-7}$ |
| Horiz. slowness at R | $p_{h,R}$, s/km | 0.15 | 0.14999992 | $-5.41 \cdot 10^{-7}$ |
| Sigma | $\sigma$, km²/s | 50 | 49.99999928 | $-1.44 \cdot 10^{-8}$ |
| Geom. spreading | $L_{GS}$, km²/s | 50 | 50.00000175 | $+3.50 \cdot 10^{-8}$ |

Table 7. Nodal locations and orientations for the conic velocity model path.

| node | location, km | | orientation | |
|---|---|---|---|---|
| | $x_1$ | $x_3$ | $r_1$ | $r_3$ |
| 0 | –5 | 0 | 0.518386 | +0.855147 |
| 1 | –4.56960 | 0.583633 | 0.661585 | +0.750320 |
| 2 | –4.04831 | 1.08774 | 0.770002 | +0.638042 |
| 3 | –3.45858 | 1.50974 | 0.851026 | +0.525123 |
| 4 | –2.81846 | 1.85051 | 0.909999 | +0.414610 |
| 5 | –2.14217 | 2.11223 | 0.951595 | +0.307357 |
| 6 | –1.44099 | 2.29724 | 0.979162 | +0.203084 |
| 7 | –0.724246 | 2.40774 | 0.994889 | +0.100979 |
| 8 | 0 | 2.44403 | 1 | 0 |
| 9 | +0.724246 | 2.40774 | 0.994889 | –0.100979 |
| 10 | +1.44099 | 2.29724 | 0.979162 | –0.203084 |
| 11 | +2.14217 | 2.11223 | 0.951595 | –0.307357 |
| 12 | +2.81846 | 1.85051 | 0.909999 | –0.414610 |
| 13 | +3.45858 | 1.50974 | 0.851026 | –0.525123 |
| 14 | +4.04831 | 1.08774 | 0.770002 | –0.638042 |
| 15 | +4.56960 | 0.583633 | 0.661585 | –0.750320 |
| 16 | +5 | 0 | 0.518386 | –0.855147 |

Table 8. Endpoint location Hessian, s/km$^2$ for conic velocity model.

The mixed $RS$ block is highlighted in yellow.

| DoF | $x_{S,1}$ | $x_{S,2}$ | $x_{S,3}$ | $x_{R,1}$ | $x_{R,2}$ | $x_{R,3}$ |
|---|---|---|---|---|---|---|
| $x_{S,1}$ | $-8.89611 \times 10^{-3}$ | 0 | $+6.21106 \times 10^{-3}$ | $+9.32373 \times 10^{-3}$ | 0 | $+5.66182 \times 10^{-3}$ |
| $x_{S,2}$ | 0 | $+2.59195 \times 10^{-2}$ | 0 | 0 | $-2.59195 \times 10^{-2}$ | 0 |
| $x_{S,3}$ | $+6.21106 \times 10^{-3}$ | 0 | $+2.89558 \times 10^{-1}$ | $-5.66182 \times 10^{-3}$ | 0 | $-3.43812 \times 10^{-3}$ |
| $x_{R,1}$ | $+9.32373 \times 10^{-3}$ | 0 | $-5.66182 \times 10^{-3}$ | $-8.89611 \times 10^{-3}$ | 0 | $-6.21106 \times 10^{-3}$ |
| $x_{R,2}$ | 0 | $-2.59195 \times 10^{-2}$ | 0 | 0 | $+2.59195 \times 10^{-2}$ | 0 |
| $x_{R,3}$ | $+5.66182 \times 10^{-3}$ | 0 | $-3.43812 \times 10^{-3}$ | $-6.21106 \times 10^{-3}$ | 0 | $+2.89558 \times 10^{-1}$ |

Table 9. Accuracy of the conic velocity model path.

| Characteristic | Notation | Exact | Numerical | Rel. error |
|---|---|---|---|---|
| Take-off angle | $\theta_a$, rad | 0.54496726 | 0.54496265 | $-8.46 \cdot 10^{-6}$ |
| Max. depth | $z_{max}$, km | 2.4440395 | 2.4440315 | $-3.30 \cdot 10^{-6}$ |
| Path arclength | $s$, km | 11.610989 | 11.61082 | $+6.09 \cdot 10^{-7}$ |
| Traveltime | $t$, s | 3.60335336584 | 3.60335336578 | $-1.71 \cdot 10^{-11}$ |
| Horiz. slowness | $p_h$, s/km | 0.25919505 | 0.25919539 | $+1.32 \cdot 10^{-6}$ |
| Sigma | $\sigma$, km$^2$/s | 38.580984 | 38.580933 | $+1.32 \cdot 10^{-6}$ |
| Geom. spreading | $L_{GS}$, km$^2$/s | 54.983109 | 54.983184 | $+1.37 \cdot 10^{-6}$ |

Table 10. Normalized characteristics of deep, shallow, and minimum offset waves in the simple, unsmooth, caustic-generating medium (computed analytically).

| Characteristic | Notation | Deep wave | Shallow wave | Min. offset wave |
|---|---|---|---|---|
| Take-off angle | $\theta_a$, rad | 0.30046834 | 0.82189389 | $\pi/6 = 30^\circ$ |
| Traveltime | $k\,t$ | 10.057414 | 10.474043 | 9.5621190 |
| Radius of circular arc | $\rho/h_{\min}$ | 0.48768030 | 0.19706494 | 0.28867513 |
| Path arclength | $s/h_{\min}$ | 2.1456728 | 1.5671691 | 1.6045998 |
| Max depth | $z_{\max}/z_h$ | 1.7929161 | 1.1217686 | $4/3$ |
| Max depth | $z_{\max}/h_{\min}$ | 0.77635544 | 0.48574007 | 0.57735027 |
| Horiz. slowness | $p_h v_a$ | 0.29596760 | 0.73243657 | $1/2$ |
| Sigma | $\sigma/(h_{\min}^2 k)$ | 0.58521636 | 0.23647793 | 0.28867513 |
| Geom. spreading | $L_{GS}/(h_{\min}^2 k)$ | 0.43509992 | 0.17581793 | 0 |
| Geom. spreading | $L_{GS}/\sigma$ | 0.74348557 | 0.74348557 | 0 |
| No. of caustics | $m_c$ | 0 | 1 | 1 |

Table 11. Normalized characteristics of deep, shallow, and minimum offset waves in the simple, smooth, caustic-generating medium (computed numerically).

| Characteristic | Notation | Deep wave | Shallow wave | Min. offset wave |
|---|---|---|---|---|
| Take-off angle | $\theta_a$, rad | 0.30058429 | 0.81940573 | 0.52252967 |
| Traveltime | $k\,t$ | 10.049339 | 10.462203 | 9.5541344 |
| Path arclength | $s/h_{\min}$ | 2.1449079 | 1.5679667 | 1.6055466 |
| Max depth | $z_{\max}/z_h$ | 1.7922001 | 1.1228080 | 1.3345532 |
| Max depth | $z_{\max}/h_{\min}$ | 0.77604541 | 0.48619012 | 0.57787850 |
| Horiz. slowness | $p_h v_a$ | 0.29616290 | 0.73075172 | 0.49893487 |
| Sigma | $\sigma/(h_{\min}^2 k)$ | 0.58483044 | 0.23702316 | 0.28929140 |
| Geom. spreading | $L_{GS}/(h_{\min}^2 k)$ | 0.43437393 | 0.17658025 | 0 |
| Geom. spreading | $L_{GS}/\sigma$ | 0.74273481 | 0.74499153 | 0 |
| No. of caustics | $m_c$ | 0 | 1 | 1 |

Table 12. Kinematic characteristics of ray paths for the gas-cloud velocity model.

| Characteristic | Notation | Ray passing through | Bypassing ray |
|---|---|---|---|
| Take-off angle | $\theta_a$, rad | 0 | +0.49973758 |
| Arrival angle | $\theta_b$, rad | 0 | −0.31131464 |
| Traveltime | $t$, s | 1.6443907 | 1.5905374 |
| Path arclength | $s$, km | 3 | 3.1554401 |
| Sigma | $\sigma$, km²/s | 6.0426767 | 6.6672085 |

Table 13. Dynamic characteristics of ray paths for the gas-cloud velocity model.

| Characteristic | Notation | Ray passing through | Bypassing ray |
|---|---|---|---|
| Geometric spreading, 2.5D | $L_{GS}$, km²/s | 5.7658207 | 9.3004414 |
| Geometric spreading, 3D | $L_{GS}$, km²/s | 5.5016503 | 0 |
| Normalized spreading, 2.5D | $L_{GS}/\sigma$ | 0.95418335 | 1.3949505 |
| Normalized spreading, 3D | $L_{GS}/\sigma$ | 0.91046578 | 0 |
| Caustic type, 2.5D | | line, $x_2$ | none |
| Caustic type, 3D | | point | line in $x_1 x_3$ |

Table 14. Nodal locations and orientations of the path,

for ellipsoidal model with constant gradient.

| node | location, km | | orientation | |
|---|---|---|---|---|
| | $x_1$ | $x_3$ | $r_1$ | $r_3$ |
| 0 | –5 | 0 | 0.760686 | +0.649119 |
| 1 | –4.46415 | 0.410307 | 0.824333 | +0.566105 |
| 2 | –3.88961 | 0.764408 | 0.875689 | +0.482876 |
| 3 | –3.28410 | 1.06247 | 0.916411 | +0.400239 |
| 4 | –2.65431 | 1.30504 | 0.947907 | +0.381546 |
| 5 | –2.00606 | 1.49280 | 0.971301 | +0.237856 |
| 6 | –1.34452 | 1.62639 | 0.987433 | +0.158038 |
| 7 | –0.674370 | 1.70632 | 0.996886 | +0.0788541 |
| 8 | 0 | 1.73293 | 1 | 0 |
| 9 | +0.674370 | 1.70632 | 0.996886 | –0.0788541 |
| 10 | +1.34452 | 1.62639 | 0.987433 | –0.158038 |
| 11 | +2.00606 | 1.49280 | 0.971301 | –0.237856 |
| 12 | +2.65431 | 1.30504 | 0.947907 | –0.381546 |
| 13 | +3.28410 | 1.06247 | 0.916411 | –0.400239 |
| 14 | +3.88961 | 0.764408 | 0.875689 | –0.482876 |
| 15 | +4.46415 | 0.410307 | 0.824333 | –0.566105 |
| 16 | +5 | 0 | 0.760686 | –0.649119 |

Table 15. Endpoint location Hessian, s/km² for ellipsoidal velocity model with constant gradient. The mixed *RS* block is highlighted in yellow.

| DoF | $x_{S,1}$ | $x_{S,2}$ | $x_{S,3}$ | $x_{R,1}$ | $x_{R,2}$ | $x_{R,3}$ |
|---|---|---|---|---|---|---|
| $x_{S,1}$ | $+8.24397 \times 10^{-3}$ | o | $+5.94385 \times 10^{-2}$ | $+9.70332 \times 10^{-3}$ | o | $+1.13744 \times 10^{-2}$ |
| $x_{S,2}$ | o | $+2.15482 \times 10^{-2}$ | o | o | $-2.15482 \times 10^{-2}$ | o |
| $x_{S,3}$ | $+5.94385 \times 10^{-2}$ | o | $+2.04881 \times 10^{-1}$ | $-1.13751 \times 10^{-2}$ | o | $-1.33341 \times 10^{-2}$ |
| $x_{R,1}$ | $+9.70332 \times 10^{-3}$ | o | $-1.13751 \times 10^{-2}$ | $-1.15219 \times 10^{-2}$ | o | $+3.31106 \times 10^{-3}$ |
| $x_{R,2}$ | o | $-2.15482 \times 10^{-2}$ | o | o | $+2.15482 \times 10^{-2}$ | o |
| $x_{R,3}$ | $+1.13744 \times 10^{-2}$ | o | $-1.33341 \times 10^{-2}$ | $+3.31106 \times 10^{-3}$ | o | $+7.26391 \times 10^{-2}$ |

Table 16. Accuracy of the ray path for ellipsoidal velocity model with constant gradient.

| Characteristic | Notation | Exact | Numerical | Rel. error |
|---|---|---|---|---|
| Take-off angle | $\theta_a$, rad | 0.8643702711 | 0.8643701148 | $-1.81 \cdot 10^{-7}$ |
| Max. depth | $z_{\max}$, km | 1.732928050 | 1.732928032 | $+1.03 \cdot 10^{-8}$ |
| Path arclength | $s$, km | 10.80186328 | 10.80186327 | $+8.26 \cdot 10^{-10}$ |
| Traveltime | $t$, s | 2.40181925390 | 2.40181925389 | $+1.84 \cdot 10^{-12}$ |
| Sigma | $\sigma$, km²/s | 51.32750192 | 51.32750182 | $-1.93 \cdot 10^{-9}$ |
| Geom. spreading | $L_{GS}$, km²/s | unknown | 46.00000094 | – |

Table 17. Conversion velocity and geometric spreading (GS)

for ellipsoidal model with constant gradient.

| Parameter | Notation | Forward path | Reverse path |
|---|---|---|---|
| Traveltime, s | $t$ | 2.401819254 | 2.401819254 |
| Arclength, km | $s$ | 10.80186327 | 10.80186327 |
| Parameter $\sigma$, km$^2$/s | $\sigma$ | 51.32750182 | 51.32750182 |
| Phase velocity at source, km/s | $v_{\text{phs},S}$ | 2.193171093 | 4.386342290 |
| Ray velocity at source, km/s | $v_{\text{ray},S}$ | 2.247775389 | 4.495550889 |
| Conversion velocity, km/s | $v_{J,S}$ | 2.304263900 | 4.608527577 |
| GS computed with Hessian, km$^2$/s | $L_{GS}$ | 46.00858689 | 46.00858689 |
| GS computed with DRT, km$^2$/s | $L_{GS}$ | 46.00000095 | 46.00000055 |
| Normalized geom. spreading at $S$ | $(L_{GS}/\sigma)_S$ | 1.037813949 | 1.037813873 |

Table 18. Nodal locations and orientations of the path,

for ellipsoidal model with varying gradient.

| node | location, km | | orientation | |
|---|---|---|---|---|
| | $x_1$ | $x_3$ | $r_1$ | $r_3$ |
| 0 | –5 | 0 | 0.555191 | +0.831723 |
| 1 | –4.55859 | 0.544416 | 0.696494 | +0.717563 |
| 2 | –4.03008 | 1.00558 | 0.802470 | +0.596692 |
| 3 | –3.43961 | 1.38203 | 0.878747 | +0.477289 |
| 4 | –2.80352 | 1.67637 | 0.931759 | +0.363077 |
| 5 | –2.13690 | 1.89277 | 0.966894 | +0.255177 |
| 6 | –1.45074 | 2.03568 | 0.988161 | +0.153422 |
| 7 | –0.753731 | 2.10919 | 0.998369 | +0.0570918 |
| 8 | –0.0528940 | 2.11679 | 0.999396 | –0.0347434 |
| 9 | +0.645783 | 2.06129 | 0.992347 | –0.123075 |
| 10 | +1.33691 | 1.94480 | 0.977937 | –0.208902 |
| 11 | +2.01531 | 1.76873 | 0.956048 | –0.293211 |
| 12 | +2.67564 | 1.53379 | 0.926232 | –0.376954 |
| 13 | +3.31201 | 1.24007 | 0.887398 | –0.461004 |
| 14 | +3.91751 | 0.887092 | 0.837754 | –0.546048 |
| 15 | +4.48371 | 0.474002 | 0.774666 | –0.632371 |
| 16 | +5 | 0 | 0.694543 | –0.719451 |

Table 19. Endpoint location Hessian, s/km² for ellipsoidal velocity model with varying gradient.

The mixed *RS* block is highlighted in yellow.

| DoF | $x_{S,1}$ | $x_{S,2}$ | $x_{S,3}$ | $x_{R,1}$ | $x_{R,2}$ | $x_{R,3}$ |
|---|---|---|---|---|---|---|
| $x_{S,1}$ | $-3.79048 \times 10^{-2}$ | 0 | $-6.33234 \times 10^{-3}$ | $+1.35200 \times 10^{-2}$ | 0 | $+1.30560 \times 10^{-2}$ |
| $x_{S,2}$ | 0 | $+1.70760 \times 10^{-2}$ | 0 | 0 | $-1.70760 \times 10^{-2}$ | 0 |
| $x_{S,3}$ | $-6.33234 \times 10^{-3}$ | 0 | $+1.22874 \times 10^{-1}$ | $-9.03626 \times 10^{-3}$ | 0 | $-8.72616 \times 10^{-3}$ |
| $x_{R,1}$ | $+1.35200 \times 10^{-2}$ | 0 | $-9.03626 \times 10^{-3}$ | $-1.71333 \times 10^{-2}$ | 0 | $+1.52979 \times 10^{-2}$ |
| $x_{R,2}$ | 0 | $-1.70760 \times 10^{-2}$ | 0 | 0 | $+1.70760 \times 10^{-2}$ | 0 |
| $x_{R,3}$ | $+1.30560 \times 10^{-2}$ | 0 | $-8.72616 \times 10^{-3}$ | $+1.52979 \times 10^{-2}$ | 0 | $+3.78955 \times 10^{-2}$ |

Table 20. Kinematic and dynamic characteristics

for ellipsoidal model with constant gradient.

| Parameter | Notation | Forward path | Reverse path |
|---|---|---|---|
| Traveltime, s | $t$ | 2.037284170 | 2.037284170 |
| Arclength, km | $s$ | 11.22021371 | 11.22021371 |
| Parameter $\sigma$, km | $\sigma$ | 62.91602468 | 62.91602468 |
| Phase velocity at source, km/s | $v_{\text{phs},S}$ | 3.388141982 | 5.643204829 |
| Ray velocity at source, km/s | $v_{\text{ray},S}$ | 3.446863109 | 5.775373550 |
| Conversion velocity, km/s | $v_{J,S}$ | 4.189296330 | 6.308886833 |
| GS computed with Hessian, km²/s | $L_{GS}$ | 51.94777288 | 51.94777288 |
| GS computed with DRT, km²/s | $L_{GS}$ | 51.93371373 | 51.93375755 |
| Normalized geom. spreading at $S$ | $(L_{GS}/\sigma)_S$ | 1.225880855 | 1.105095449 |

# LIST OF FIGURES

Figure 1. Examples 1a and 2: Rays in a constant velocity gradient model and a conic velocity model: a) circular ray path in medium with a constant velocity gradient (linear velocity model, solid line) and an elliptic ray path in medium with a conic velocity model (dashed line), b) geometric spreading vs. arclength, c) normalized geometric spreading $L_{GS}/\sigma$. Different segment colors correspond to eight finite elements.

Figure 2. Example 1b: Rays in tilted constant velocity gradient model: a) circular ray path, b) geometric spreading for forward and reverse paths shown by solid and dashed lines, respectively, c) normalized geometric spreading.

Figure 3. Example 2: Conic velocity model: a) velocity profile, b) velocity gradient profile.

Figure 4. Example 3: Simple, smooth, caustic-generating model: a) velocity profile, b) velocity gradient, c) second derivative of the velocity.

Figure 5. Example 3: Pre-critical ray in the simple, smooth, caustic-generating medium: a) ray trajectory; dashed gray line is the initial guess, b) geometric spreading, c) normalized geometric spreading

Figure 6. Example 3: Post-critical ray in the simple, smooth, caustic-generating medium: a) ray trajectory; dashed gray line is the initial guess, b) geometric spreading, c) normalized geometric spreading.

Figure 7. Example 3: Critical ray in the simple, smooth, caustic-generating medium: a) ray trajectory; dashed gray line is the initial guess, b) geometric spreading, c) normalized geometric spreading.

Figure 8. Example 4: 2.5D gas-cloud velocity model with the constant vertical (along $x_3$) gradient half-space and cylindrical anomaly: a) velocity distribution, b) absolute value of the velocity gradient.

Figure 9. Example 4a: Rays in 2.5D gas-cloud model: a) ray trajectories, b) geometric spreading, c) normalized geometric spreading.

Figure 10. Example 4b: Rays in 3D gas-cloud model: a) geometric spreading, b) normalized geometric spreading.

Figure 11. Example 5a: Rays in ellipsoidal model with tilted constant velocity gradient background: a) elliptic ray path, b) geometric spreading for forward and reverse paths shown by solid and dashed lines, respectively, c) normalized geometric spreading.

Figure 12. Example 5b: Rays in ellipsoidal model with tilted varying velocity gradient background: a) asymmetric ray path, b) geometric spreading for forward and reverse paths shown by solid and dashed lines, respectively, c) normalized geometric spreading.

Figure 13. Example 3: Ray trajectory in the simple, unsmooth, caustic-generating medium. Red line is the ray path, blue line is the interface between the layer and the half-space, and black line is the earth's surface.

Figure 14. Example 3: Rays in a medium with simple, unsmooth, caustic-generating velocity model: a) normalized offset $h/h_{min}$ vs. take-off angle, b) ray trajectories with the caustic (red line) and with no caustic (blue line) corresponding to the same offset $h = 1.2 h_{min}$ (multi-arrival); gray line is the interface between the constant velocity layer and the constant velocity gradient half-space, c) ray with no caustics and its paraxial rays, d) ray with a caustic and its paraxial rays, e) intersection of the central ray (with a caustic) and its paraxial rays, f) zoom demonstrating that central and paraxial rays intersect at different points, g) minimum offset ray and its two paraxial rays, h) intersection of minimum-offset ray with its paraxial ray of higher take-off angle.

Figure 15. Stationary reflection path in the simple caustic-generating model.

Figure 16. Example 3: Reflection rays in the simple caustic-generating model: a) two-way traveltime vs. reflector depth, b) derivative of the traveltime wrt the reflector depth, c) incidence/reflection angle vs. reflector depth.

Figure 17. Example 3: Path of paraxial ray for the simple, unsmooth, caustic-generating velocity model, with take-off angle $\omega = \theta_c + \gamma$, split into four intervals.

Figure 18. Jacobian vs. traveltime for the simple, unsmooth, caustic-generating velocity model: a) caustic-free pre-critical ray, with the offset $h = 1.2 h_{min}$, b) post-critical ray with a caustic, with the same offset $h = 1.2 h_{min}$, c) ray with the critical take-off angle and the minimum offset $h_{min}$, allowing the diving ray, where a caustic occurs at the destination point.

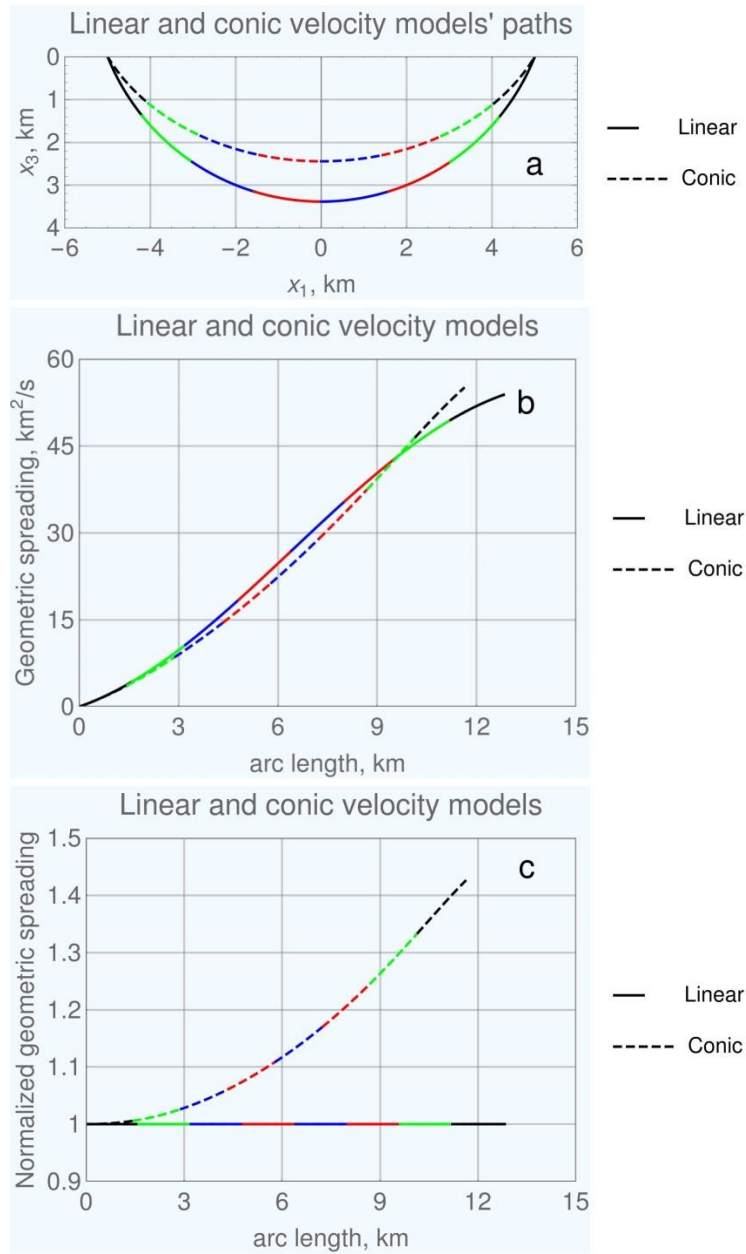

Figure 1. Examples 1a and 2: Rays in a constant velocity gradient model and a conic velocity model: a) circular ray path in medium with a constant velocity gradient (linear velocity model, solid line) and an elliptic ray path in medium with a conic velocity model (dashed line), b) geometric spreading vs. arclength, c) normalized geometric spreading $L_{GS}/\sigma$. Different segment colors correspond to eight finite elements.

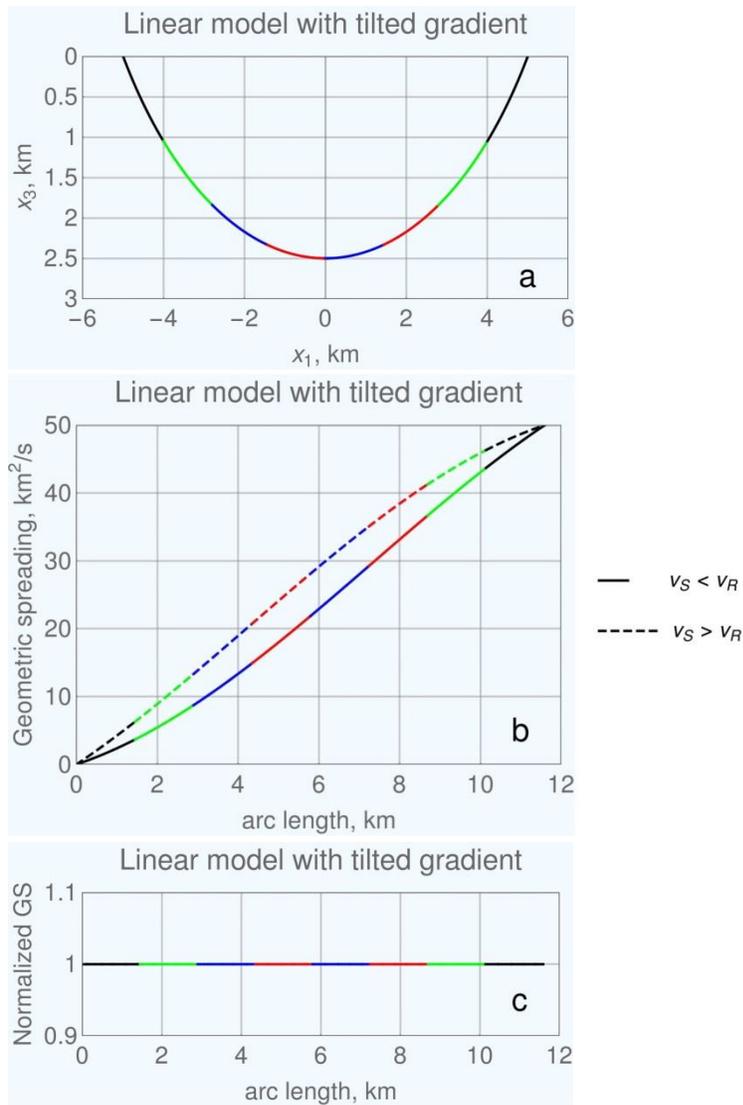

Figure 2. Example 1b: Rays in tilted constant velocity gradient model: a) circular ray path, b) geometric spreading for forward and reverse paths shown by solid and dashed lines, respectively, c) normalized geometric spreading.

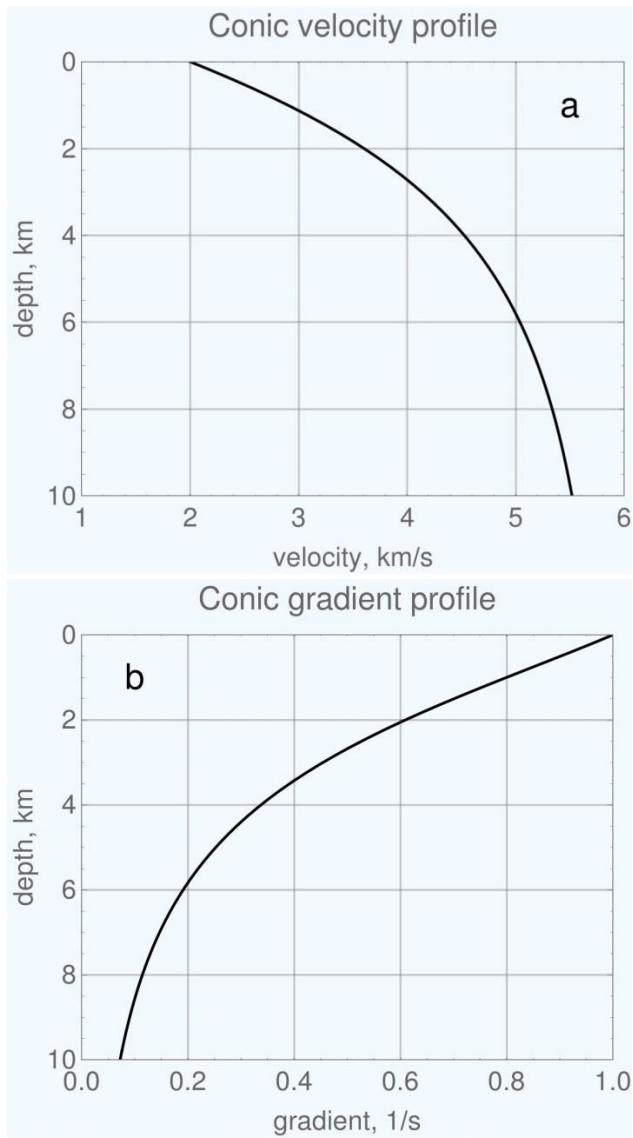

Figure 3. Example 2: Conic velocity model: a) velocity profile, b) gradient profile.

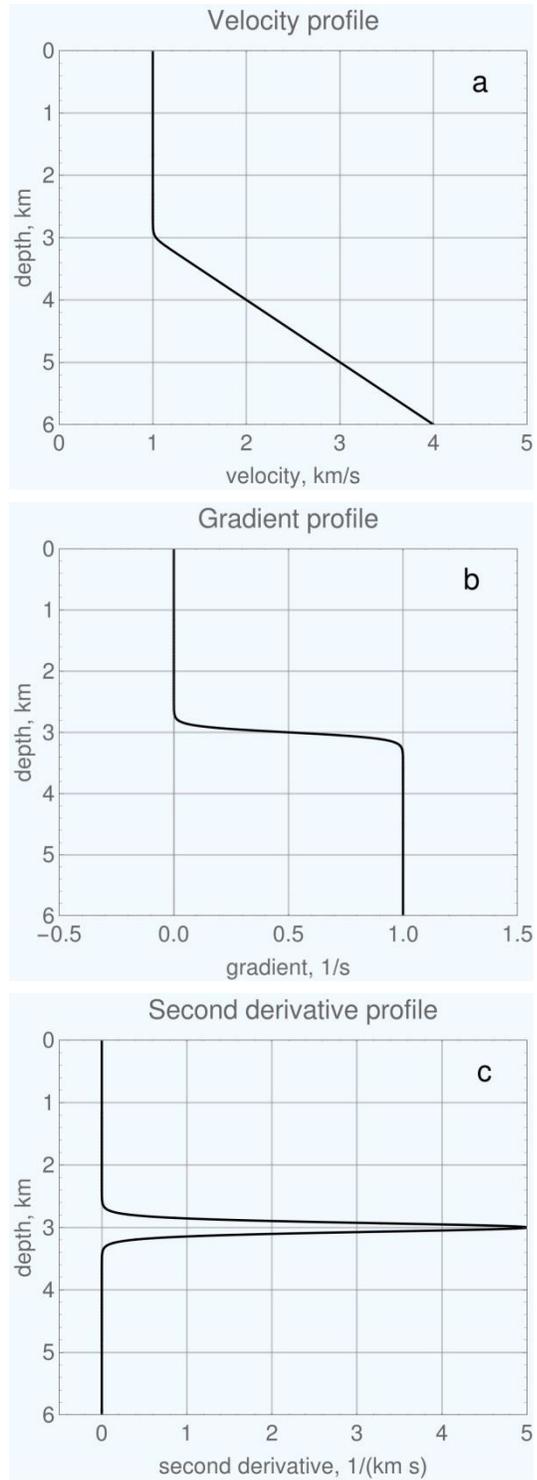

Figure 4. Example 3: Simple, smooth, caustic-generating model: a) velocity profile, b) velocity gradient, c) second derivative of the velocity.

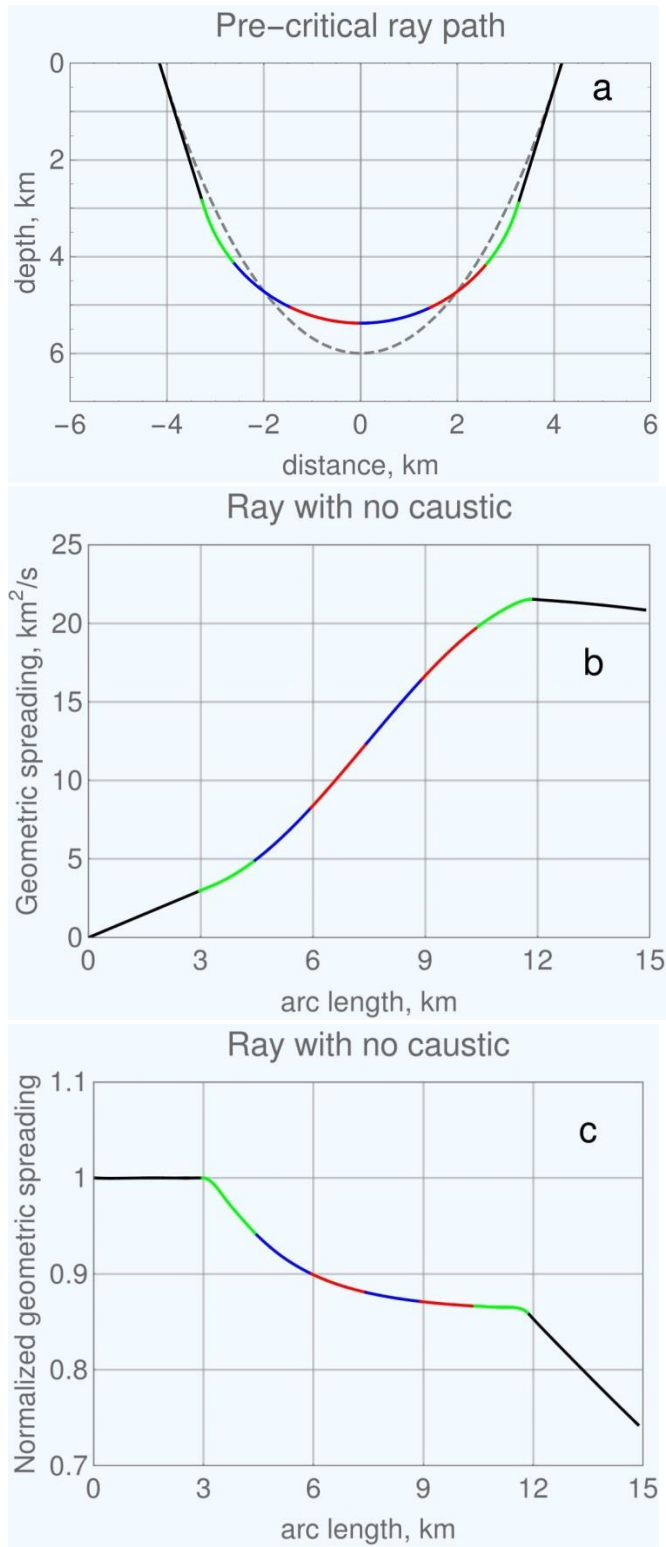

Figure 5. Example 3: Pre-critical ray in the simple, smooth, caustic-generating medium: a) ray trajectory; dashed gray line is the initial guess, b) geometric spreading, c) normalized geometric spreading.

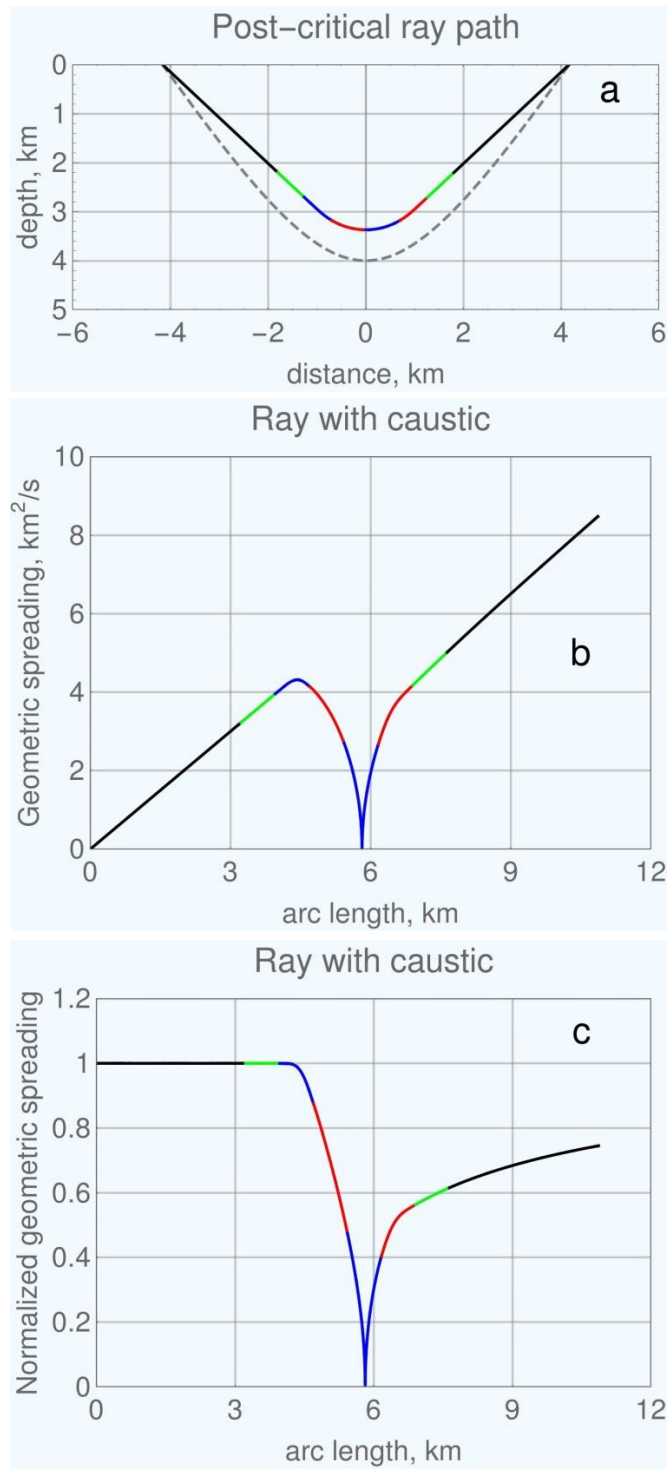

Figure 6. Example 3: Post-critical ray in the simple, smooth, caustic-generating medium: a) ray trajectory; dashed gray line is the initial guess, b) geometric spreading, c) normalized geometric spreading.

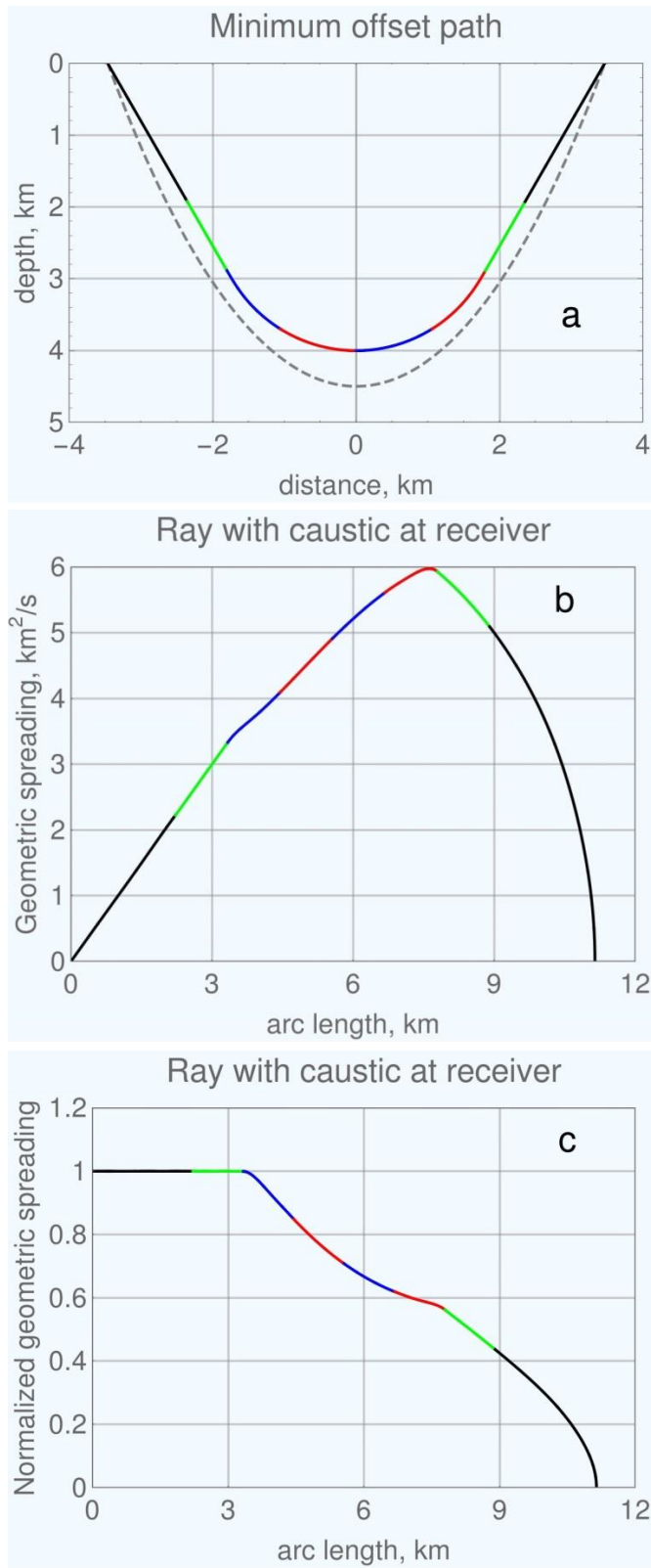

Figure 7. Example 3: Critical ray in the simple, smooth, caustic-generating medium: a) ray trajectory; dashed gray line is the initial guess, b) geometric spreading, c) normalized geometric spreading.

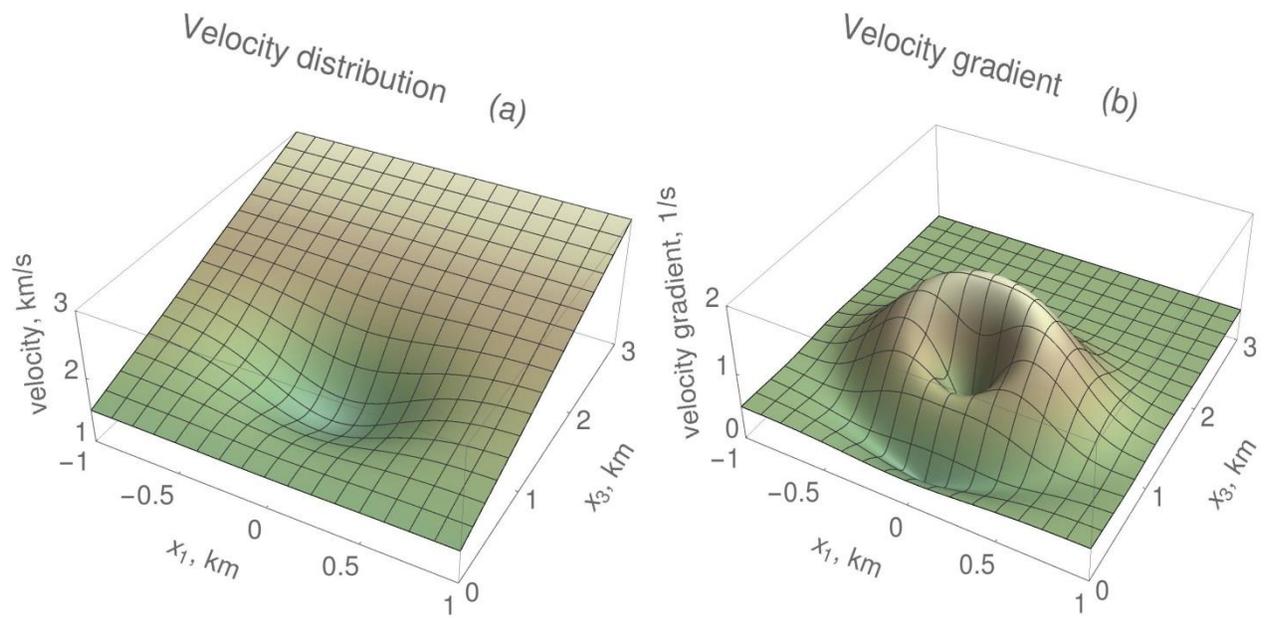

Figure 8. Example 4a. 2.5D gas-cloud velocity model with the constant vertical (along $x_3$) gradient half-space and cylindrical anomaly: a) velocity distribution, b) absolute value of the velocity gradient.

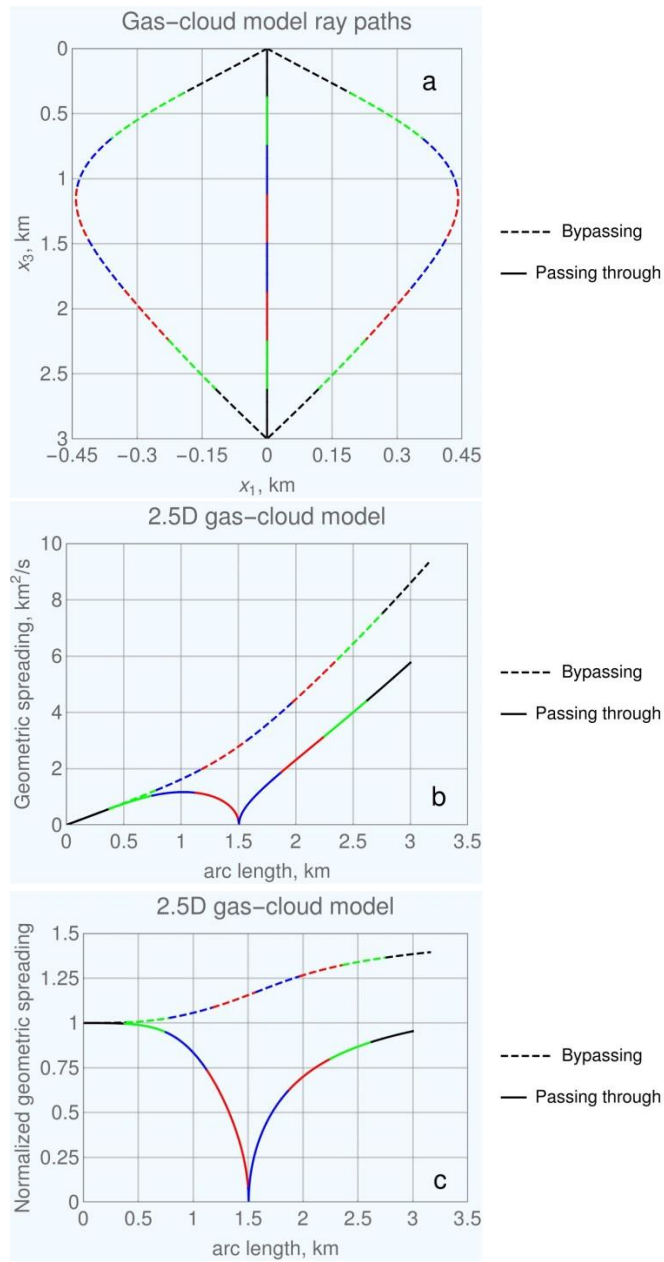

Figure 9. Example 4a: Rays in a 2.5D gas-cloud model: a) ray trajectories, b) geometric spreading, c) normalized geometric spreading.

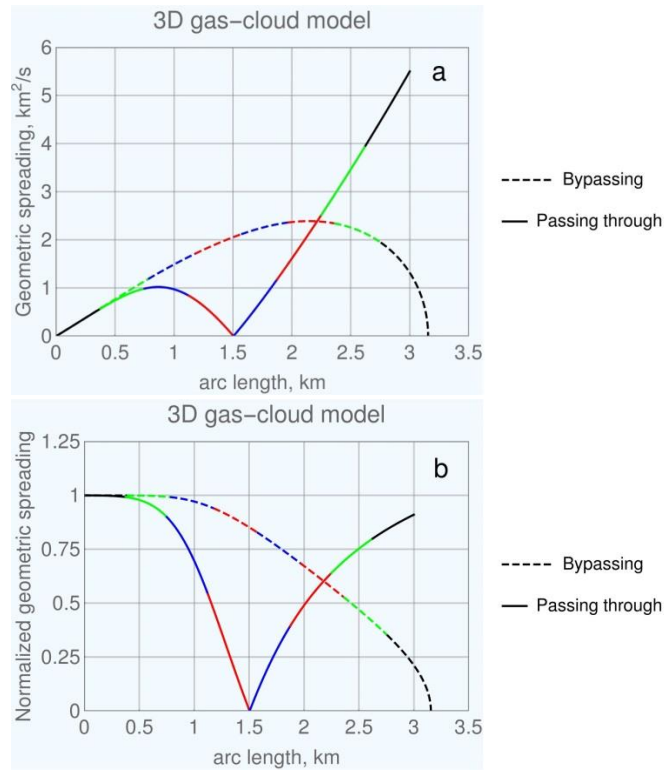

Figure 10. Example 4b: Rays in a 3D gas-cloud model: a) geometric spreading, b) normalized geometric spreading.

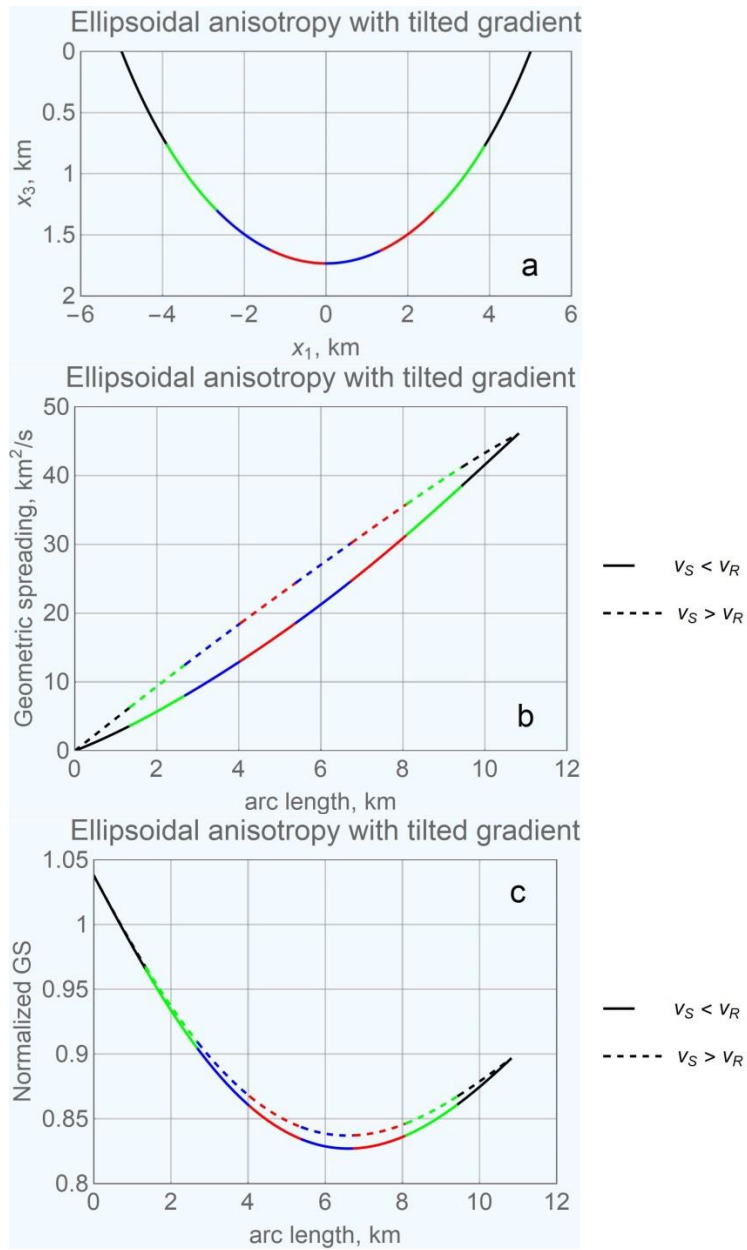

Figure 11. Example 5a: Rays in ellipsoidal model with tilted constant velocity gradient background: a) elliptic ray path, b) geometric spreading for forward and reverse paths shown by solid and dashed lines, respectively, c) normalized geometric spreading.

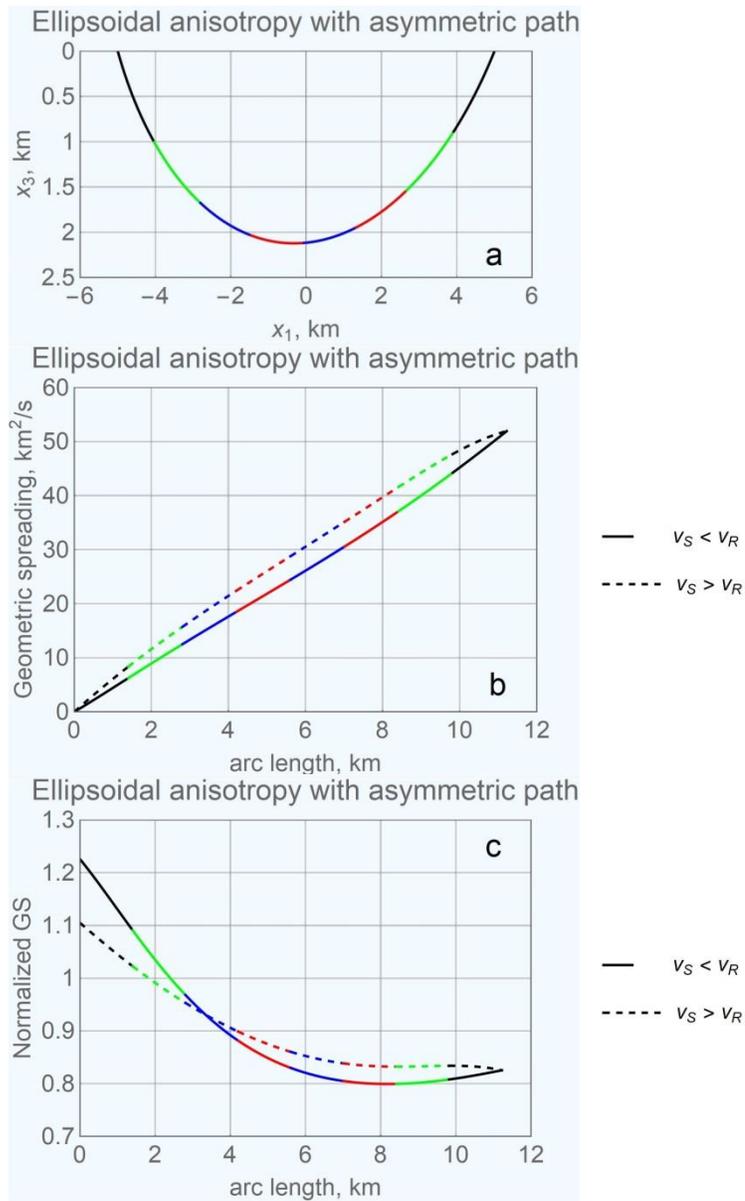

Figure 12. Example 5b: Rays in ellipsoidal model with tilted varying velocity gradient background: a) asymmetric ray path, b) geometric spreading for forward and reverse paths shown by solid and dashed lines, respectively, c) normalized geometric spreading.

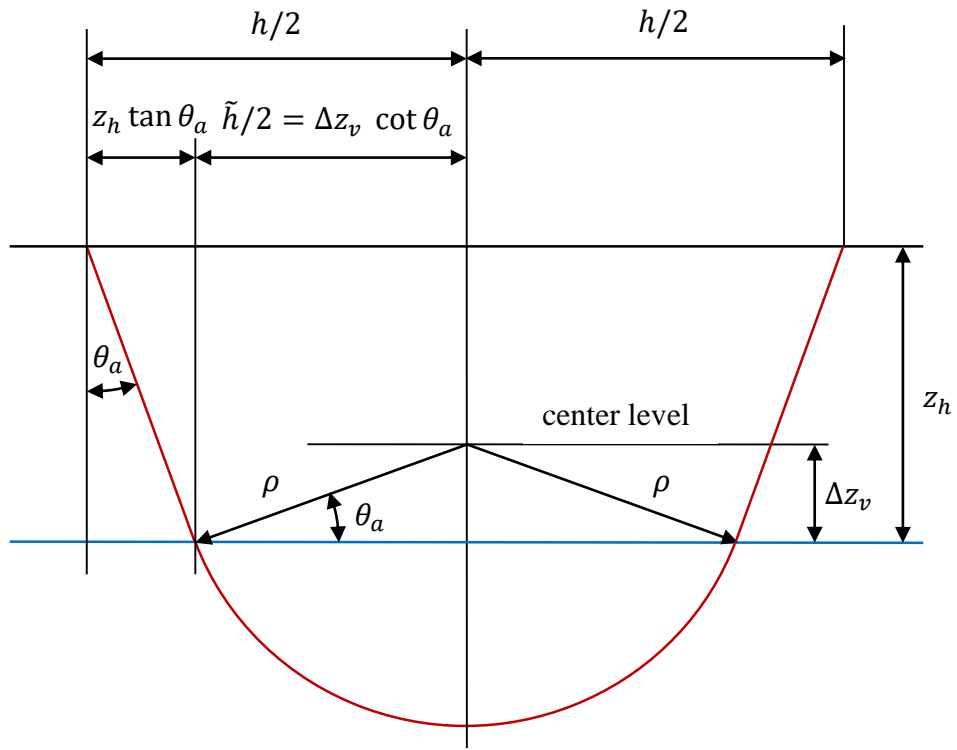

Figure 13. Example 3: Ray trajectory in the simple, unsmooth, caustic-generating medium. Red line is the ray path, blue line is the interface between the layer and the half-space, and black line is the earth's surface.

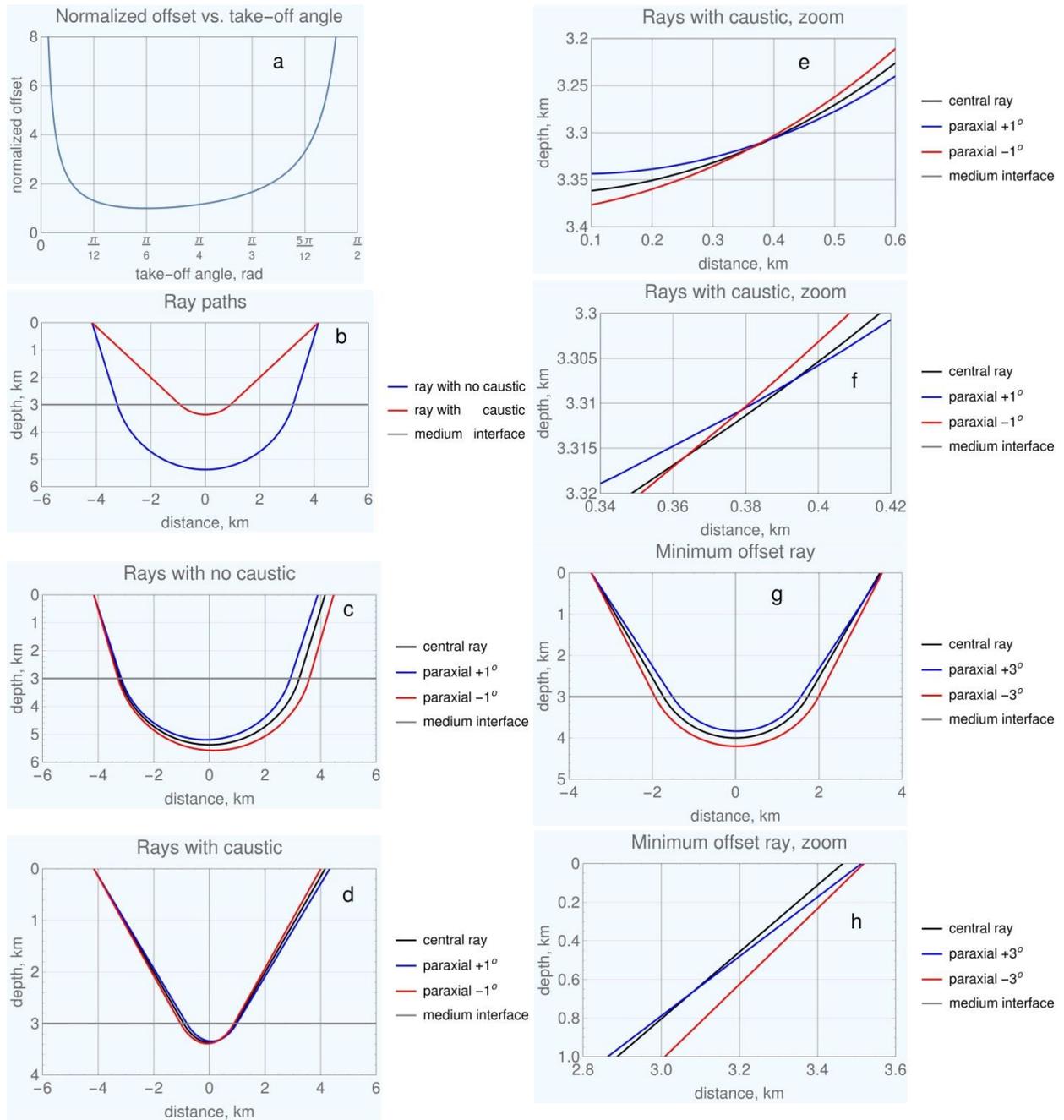

Figure 14. Example 3: Rays in a medium with the simple, unsmooth, caustic-generating velocity model: a) normalized offset $h/h_{\min}$ vs. take-off angle, b) ray trajectories with the caustic (red

line) and with no caustic (blue line) corresponding to the same offset $h = 1.2h_{min}$ (multi-arrival); gray line is the interface between the constant velocity layer and the constant velocity gradient half- space, c) ray with no caustics and its paraxial rays, d) ray with caustic and its paraxial rays, e) intersection of the central ray (with a caustic) and its paraxial rays, f) zoom demonstrating that central and paraxial rays intersect at different points, g) minimum offset ray and its two paraxial rays, h) intersection of minimum-offset ray with its paraxial ray of higher take-off angle.

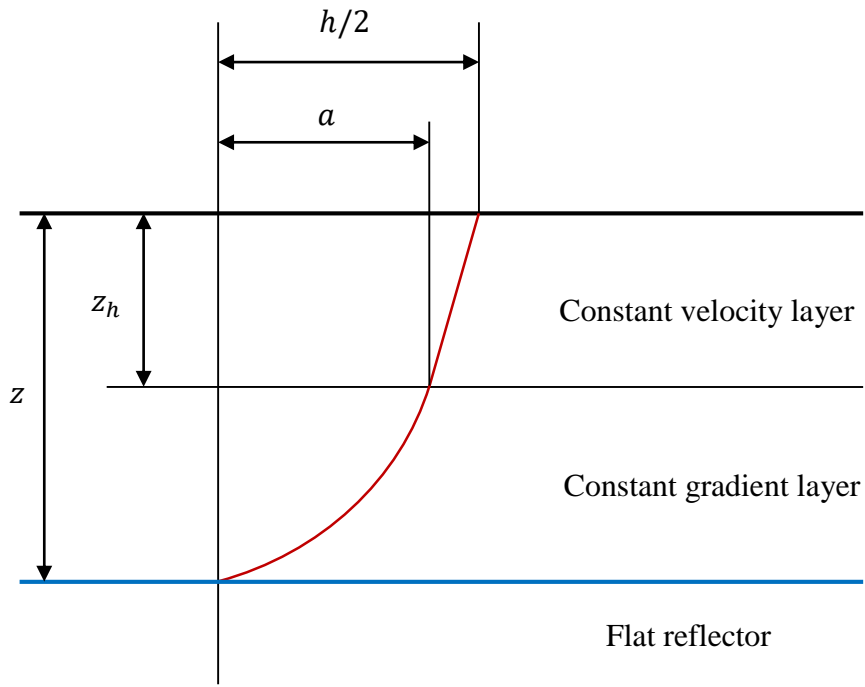

Figure 15. Stationary reflection path in the simple, caustic-generating model.

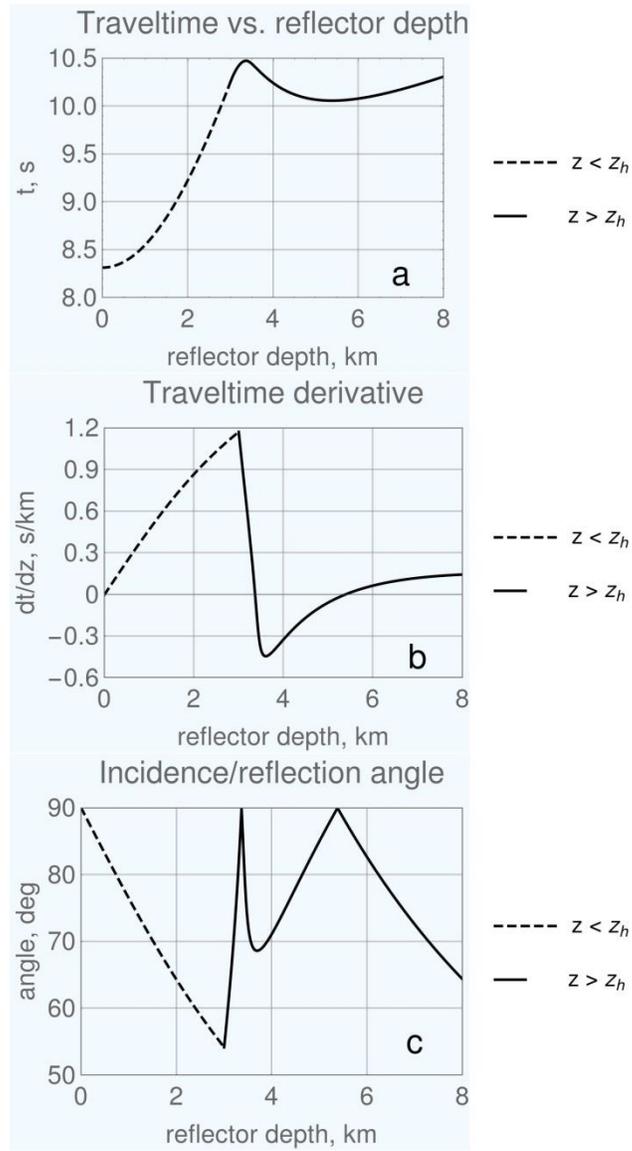

Figure 16. Example 3: Reflection rays in the simple, caustic-generating model: a) two-way traveltime vs. reflector depth, b) derivative of the traveltime wrt the reflector depth, c) incidence/reflection angle vs. reflector depth.

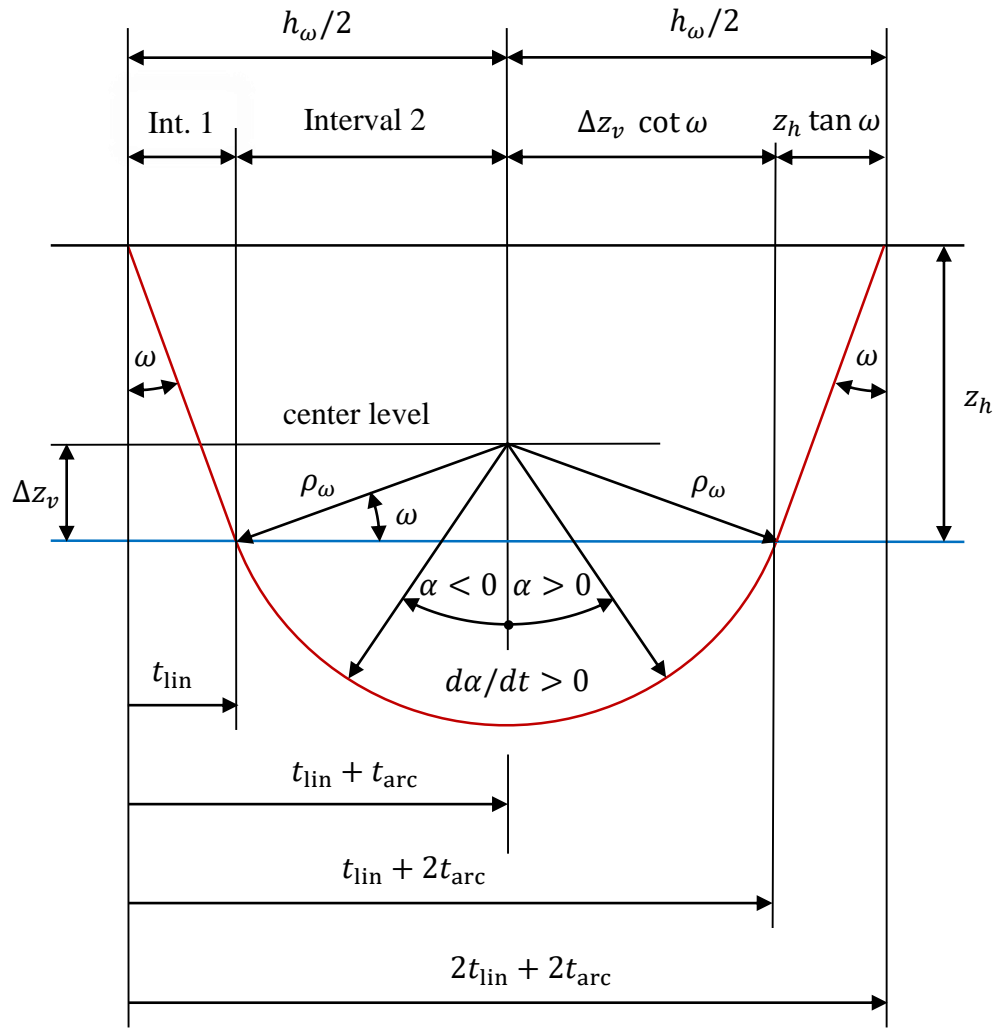

Figure 17. Example 3: Path of paraxial ray for the simple, unsmooth, caustic-generating velocity model, with take-off angle $\omega = \theta_c + \gamma$, split into four intervals.

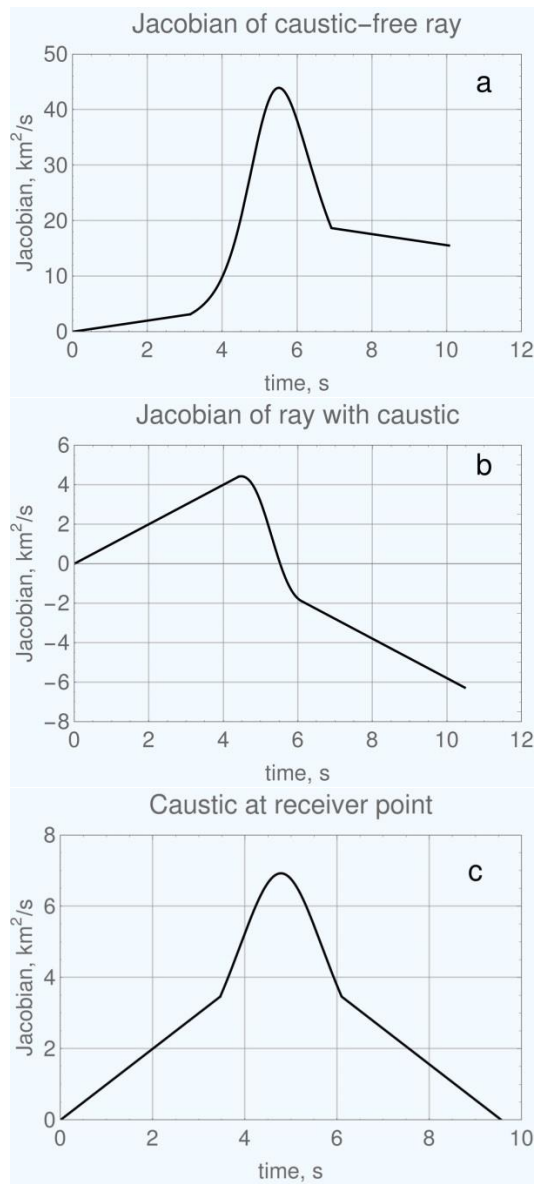

Figure 18. Example 3: Jacobian vs. traveltime for the simple, unsmooth, caustic-generating velocity model: a) caustic-free pre-critical ray, with the offset $h = 1.2 h_{min}$, b) post-critical ray with a caustic, with the same offset $h = 1.2 h_{min}$, c) ray with the critical take-off angle and minimum offset $h_{min}$ allowing the diving ray, where a caustic occurs at the destination point.